\documentstyle[ssi,epsf]{article}
\newcommand{\ewxy}[2]{\setlength{\epsfxsize}{#2}\epsfbox[10 60 640 570]{#1}}

\newcommand{\bee}{\begin{equation}}
\newcommand{\ee}{\end{equation}}
\newcommand{\beea}{\begin{eqnarray}}
\newcommand{\eea}{\end{eqnarray}}
\newcommand{\MC}{M_{hc}}
\newcommand{\MHS}{M_{hs}}
\newcommand{\MS}{M_{bs}}
\newcommand{\MB}{M_{bd}}
\newcommand{\rme}{{\rm e}}
\newcommand{\cO}{{\cal O}}
\newcommand{\Dd}[1]{\mbox{
  \parbox[b]{0cm}{$D$}\raisebox{1.7ex}{$\leftrightarrow$}$_{\!#1}$}}

\begin{document}
\title{ LATTICE GAUGE THEORY FOR QCD}
\author{
Thomas DeGrand \\
Department of Physics\\
 University of Colorado, Boulder CO 80309-390\\ [0.4cm]
}

\maketitle
\begin{abstract}
These lectures provide an introduction to lattice methods for
nonperturbative studies of Quantum Chromodynamics. 
Lecture 1 (Ch. 2): Basic techniques for QCD and results for hadron
spectroscopy using the simplest discretizations; lecture 2 (Ch. 3):
 ``improved actions''--what they are and how well they work;
lecture 3 (Ch. 4): SLAC physics from the lattice:
structure functions, the mass of the glueball, heavy quarks and
  $\alpha_s(M_Z)$, and $B-\bar B$ mixing.
\end{abstract}

\section{Introduction}

The lattice \cite{STANDARD} version of QCD was 
invented by Wilson \cite{KEN} in 1974.
It has been a fruitful source of qualitative and 
quantitative information about QCD, the latter
especially in the years since Creutz, Jacobs, and Rebbi \cite{CJR}
performed the first numerical simulations of a lattice gauge theory.
Lattice methods are presently
the only way to compute masses  and  matrix elements
in the strong interactions beginning with the Lagrangian of QCD and
including no additional parameters. In the past few years the
quality of many lattice predictions has become very high,
and they are beginning to have a large impact in the wide arena
of ``testing the standard model.''
My goal in these lectures is to give enough of an overview of
the subject that an outsider will be able to make an intelligent
appraisal of a lattice calculation when  s/he encounters one later on.

The first lecture will describe why one puts QCD on a lattice, and how it
is done. This is a long story with a lot of parts, but at the end I will
 show you ``standard'' lattice results for light hadron spectroscopy.
The main problem with these calculations is that they are so unwieldy:
to get continuum-like numbers requires very large scale numerical simulations
on supercomputers, which can take years to complete
 (sort of like the high energy experiments themselves, except that
 we do not have to stack lead bricks).
We would like to reduce the computation burden of our calculations.
In Lecture Two I will describe some of the different philosophies
and techniques which are currently being used to invent ``improved
actions.''  Some of these methods actually work: some QCD problems can be
studied on very large work stations.
Finally, in Lecture Three I will give a survey of recent lattice
results for matrix elements, using physics done at SLAC as my unifying theme.

\section{Gauge Field Basics }
\subsection{Beginnings}

The lattice is a cutoff which regularizes the
ultraviolet divergences of quantum field theories. As with any regulator,
it must be removed after renormalization. Contact with experiment only
exists in the continuum limit, when the lattice spacing is taken to zero.

We are drawn to lattice methods by our desire to study nonperturbative
phenomena.  Older regularization schemes are tied closely to perturbative
expansions: one calculates a process to some order in a coupling constant;
divergences are removed order by order in perturbation theory.  The lattice,
however, is a nonperturbative cutoff. Before a calculation begins,
all wavelengths less than a lattice spacing are removed.
Generally one cannot carry out analytical studies of a field theory
for physically interesting parameter values.  However, lattice
techniques lend themselves naturally to implementation on digital
computers, and one can perform more-or-less realistic simulations of
quantum field theories, revealing their nonperturbative structure, on
present day computers.  I think it is fair to say that little of the
 quantitative results about QCD
which have been obtained in the last decade, could  have been gotten 
without the use of numerical methods.

On the lattice we sacrifice Lorentz invariance but preserve all internal
symmetries, including local gauge 
invariance.  This preservation is important for nonperturbative physics. 
 For example, gauge invariance is a property of the continuum theory 
which is nonperturbative, so maintaining it as we pass to the lattice 
means that all of its consequences (including current conservation and 
renormalizability) will be preserved.

It is very easy to write down an action for
 scalar fields regulated by a  lattice.  One just replaces the
space-time coordinate $x_\mu$ by a set of integers $n_\mu$ ($x_\mu=an_\mu$,
where $a$ is the lattice spacing). 
Field variables $\phi(x)$ are defined on sites $\phi(x_n) \equiv \phi_n$,
 The action, an integral over
the Lagrangian, is replaced by a sum over sites
\bee
\beta S = \int d^4x {\cal L} \rightarrow a^4 \sum_n 
{\cal L}(\phi_n) .\label{2.1}
\ee
and the generating functional for Euclidean Green's functions is
replaced by an ordinary integral over the lattice fields
\bee
Z = \int (\prod_n d \phi_n ) e^{-\beta S}. \label{2.2} 
\ee
Gauge fields are a little more complicated. They carry a
space-time index $\mu$ in addition to an internal symmetry index $a$
($A_\mu^a(x))$ and are associated with a path in space $x_\mu(s)$: a
particle traversing a contour in space picks up a phase factor
\bee
\psi \rightarrow P(\exp \ ig \int_s dx_\mu A_\mu) \psi
\ee
\bee
 \equiv U(s)\psi(x). \label{2.3}
\ee
$P$ is a path-ordering factor analogous to the time-ordering
operator in ordinary quantum mechanics. Under a gauge transformation $g$,
$U(s)$ is rotated at each end:
\bee
U(s) \rightarrow g^{-1}(x_\mu(s))U(s)g(x_\mu(0)). \label{2.4} 
\ee
These considerations led Wilson \cite{KEN} to formulate gauge fields
on a space-time lattice, as follows:

The fundamental variables are elements of the gauge  group $G$ which live
on the links of a four-dimensional lattice, connecting $x$ and $x+ \mu$:
$U_\mu(x)$, with $U_\mu(x+\mu)^\dagger = U_\mu(x)$
 \bee
U_\mu(n)= \exp (igaT^aA^a_\mu(n))  \label{2.5}
\ee
for $SU(N)$.
($g$ is the coupling, $a$ the lattice spacing, $A_\mu$ the vector potential, 
and $T^a$ is a group generator).

Under a gauge transformation link variables transform as 
\bee
U_\mu (x) \rightarrow V(x) U_\mu (x) V(x+ \hat \mu)^\dagger  \label{2.10}
\ee
and site variables as 
\bee
\psi(x) \rightarrow V(x) \psi(x) \label{2.11}
\ee
so the only gauge invariant operators we can use as order parameters are 
 matter fields connected by  oriented ``strings" of U's (Fig. 1a)
\bee
\bar \psi(x_1) U _\mu(x_1)U_\mu(x_1+\hat \mu)\ldots  \psi (x_2)  \label{2.13}
\ee
or closed  oriented loops of U's (Fig. 1b)
\bee
{\rm Tr} \ldots U _\mu(x)U_\mu(x+\hat \mu)\ldots \rightarrow
{\rm Tr} \ldots U_\mu(x)V^\dagger (x+ \hat \mu)V(x+ \hat \mu)
U_\mu(x+\hat \mu)\ldots  .\label{2.12}
\ee

An action is specified by recalling that the classical Yang-Mills
action involves the curl of $A_\mu$, $F_{\mu\nu}$.
Thus a lattice action ought to involve a product of
$U_\mu$'s around some closed contour. There is enormous
arbitrariness at this point. We are trying to write down a bare action.
So far, the only requirement we  want to
impose is gauge invariance, and that will be automatically satisfied
for actions built of powers of traces of U's around closed loops,
with arbitrary coupling constants.
If we assume that the gauge fields are smooth, we can expand the link
variables in a power series in  $gaA_\mu's$. For almost any closed loop, the
leading term in the expansion will be proportional to $F_{\mu\nu}^2$.
We might want our action to have the same normalization as the continuum action.
This would provide one constraint among the lattice coupling constants.

 The simplest  contour has a perimeter of four links. In $SU(N)$
\bee
\beta S={{2N} \over {g^2}}\sum_n \sum_{\mu>\nu}{\rm  Re \ Tr \ }
\big( 1 - U_\mu(n)U_\nu(n+\hat\mu)
U^\dagger  _\mu(n+\hat\nu) U^\dagger  _\nu(n) \big).  \label{2.6}
\ee
This action is called the ``plaquette action'' or the
``Wilson action'' after its inventor.
The lattice parameter $\beta=2N/g^2$ is often written instead of
$g^2=4\pi\alpha_s$.

Let us see how this action reduces to the standard continuum action.
Specializing to the U(1) gauge group, and slightly redefining the coupling,
\bee
S= {1 \over {g^2}}\sum_n \sum_{\mu  > \nu} 
{\rm Re \ }(1 - \exp(iga[A_\mu(n) +A_\nu(n + \hat \mu)
-A_\mu(x+\hat\nu)-A_\nu(n)])).\label{2.7}
\ee
The naive continuum limit is taken by assuming that the lattice spacing 
$a$ is small, and Taylor expanding
\bee
A_\mu(n+\hat\nu) = A_\mu(n) + a \partial_\nu A_\mu(n) + \ldots \label{2.8}
\ee
so the action becomes
\bee
\beta S = {1 \over {g^2}}\sum_n \sum_{\mu > \nu}
1-{\rm Re \ }( \exp(iga[a(\partial_\nu A_\mu -\partial_\mu A_\nu) + O(a^2)])) 
\ee
\bee
={1 \over {4g^2}}a^4\sum_n \sum_{\mu\nu} g^2F_{\mu\nu}^2 + \ldots 
\ee
\bee
= {1 \over 4}\int d^4 x F_{\mu\nu}^2  \\ \nonumber \label{2.9}
\ee
transforming the sum on sites back to an integral.

\begin{figure}
\centerline{\ewxy{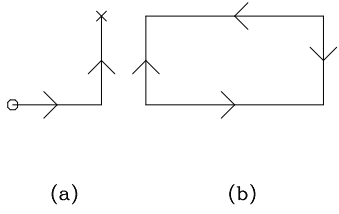}{80mm}
}
\caption{ Gauge invariant observables are
either  (a) ordered chains (``strings'')
of links connecting quarks and antiquarks or
(b) closed loops of link variables.}
\label{fig:figone}
\end{figure}

\subsection{Relativistic Fermions on the Lattice}

Defining fermions on the lattice presents a  new problem: doubling. The naive
procedure of discretizing the continuum fermion action results in a lattice model
with many more low energy modes than one originally anticipated. 
 Let's illustrate this with free field theory.

The free Euclidean fermion action in the continuum  is
\bee
S = \int d^4 x [ \bar \psi(x) \gamma_\mu \partial_\mu \psi(x) + m \bar \psi(x)
\psi(x)  ] .\label{2.23}
\ee
One obtains the so-called naive lattice formulation by replacing the derivatives
by symmetric differences: we explicitly introduce the lattice spacing $a$
in the denominator and
write
\bee
S_L^{naive} = \sum_{n,\mu} \bar \psi_n {\gamma_\mu \over {2a}}
(\psi_{n+\mu} - \psi_{n-\mu}) + m \sum_n \bar \psi_n \psi_n . \label{2.24}
\ee
The propagator is:
\bee
G(p) = (i \gamma_\mu \sin p_\mu  a + ma)^{-1} 
= {{-i \gamma_\mu \sin p_\mu a + ma}\over{\sum_\mu \sin^2 p_\mu a + m^2 a^2}} 
\label{2.25}
\ee
We identify  the physical spectrum through the poles in the
propagator, at $p_0=iE$:
\bee
\sinh^2 Ea = \sum_j \sin^2 p_j a + m^2a^2
\ee
The lowest energy solutions are the expected ones
 at $p= (0,0,0)$, $E \simeq \pm m$, but  there are 
other degenerate
ones, at $p = (\pi,0,0)$, $(0,\pi,0,)$, \dots $(\pi,\pi,\pi)$.
This is a model for eight light fermions, not one.

\noindent
(a) Wilson Fermions

There are two ways to deal with the doublers.  The first way is to alter
 the dispersion relation so that it has only one low energy solution.  The
other solutions are forced to $E \simeq 1/a$ and become very heavy as $a$
is taken to zero.  The simplest version of this solution (and almost
the only one seen in the literature until recently) 
is due to Wilson: add a second-derivative-like term
\bee
S^W = -{r \over {2a}}\sum_{n,\mu}\bar \psi_n(\psi_{n+\mu} -2 \psi_n
+\psi_{n-\mu} ) \label{2.26}
\ee
to $S^{naive}$.  The parameter $r$ must lie between 0 and 1; $r=1$ is 
almost always used and  ``$r=1$'' is implied when one
speaks of using ``Wilson fermions.''  The propagator is
\bee
G(p) = {{-i \gamma_\mu \sin p_\mu a + m a -r \sum_\mu (\cos p_\mu a -1)} \over
{\sum_\mu \sin^2 p_\mu a + (m  a-r \sum_\mu(\cos p_\mu a -1))^2}} .
\label{2.27}
\ee
It has one pair of poles
 at $p_\mu \simeq (\pm im,0,0,0)$, plus other poles at $p \simeq r/a$.
In the continuum these states become infinitely massive and decouple
(although decoupling is not  trivial to prove).

With Wilson fermions it is conventional not to use not the mass but the
``hopping parameter'' $\kappa = {1 \over 2}(m a + 4r)^{-1}$, and to
rescale the fields $\psi \rightarrow \sqrt{2 \kappa} \psi$.  The action for an
interacting theory is then written
\bee
S = \sum_n \bar \psi_n \psi_n -  \kappa \sum_{n \mu}(\bar \psi_n
(r - \gamma_\mu) U_\mu(n) \psi_{n+ \mu} + 
\bar \psi_n(r + \gamma_\mu) U_\mu^\dagger \psi_{n - \mu} ). \label{2.28}
\ee
Wilson fermions are closest to the continuum 
formulation-- there is a four component spinor on every lattice site for
every color and/or flavor of quark.  Constructing currents and states
is  just like in the continuum. 

However, the Wilson term explicitly breaks chiral symmetry.  
This has the 
consequence that the zero bare
 quark mass limit is not respected by interactions;
the quark mass is additively renormalized.    The value of $\kappa_c$,
the value of the  
hopping parameter at which the pion mass vanishes, is
not known a priori before beginning a simulation; it must be computed.
This is done in a simulation involving Wilson fermions 
 by varying $\kappa$ and watching  the pion mass 
extrapolate quadratically to zero as
$m_\pi^2  \simeq \kappa_c - \kappa$ ($\kappa_c - \kappa $ is proportional
to the quark mass for small $m_q$).
For the lattice person, this is unpleasant since preliminary 
calculations are required to find ``interesting'' $\kappa$ values. For the
outsider trying to read lattice papers, it is unpleasant because
the graphs in the lattice paper typically list $\kappa$, and not quark (or
pion) mass, so the reader does not know ``where'' the simulation was done.
Note also that the relation between $\kappa$ and physical
 mass changes with lattice
coupling $\beta$.

\noindent
(b) Staggered or Kogut-Susskind Fermions

In this formulation one reduces the number of fermion flavors by using
one component ``staggered'' fermion fields rather than four component Dirac
spinors.  The Dirac spinors are constructed by combining staggered fields
on different lattice sites.
Staggered fermions preserve an explicit chiral symmetry as $m_q \rightarrow 0$
even for finite lattice spacing, as long as all four flavors are degenerate.
They are preferred over Wilson fermions in situations in which
the chiral properties of the fermions dominate the dynamics--for
example, in studying the chiral restoration/deconfinement transition
at high temperature.  They also present a computationally less intense
situation from the point of view of numerics than Wilson fermions, for the
trivial reason that there are less variables.
  However, flavor symmetry and translational
symmetry are all mixed together.  Construction of meson and baryon states
(especially the $\Delta$) is more complicated than for Wilson 
fermions \cite{GANDSMIT}.

\subsection{Enter the Computer}
A ``generic'' Monte Carlo simulation in QCD
breaks up naturally into two parts. In the ``configuration generation''
phase one constructs an ensemble of states with the appropriate
Boltzmann weighting: we compute observables simply by
averaging $N$ measurements using the field
variables $\phi^{(i)}$ appropriate to the sample
\bee
\langle{\Gamma}\rangle \simeq \bar \Gamma
\equiv {1 \over N}\sum_{i=1}^N\Gamma[\phi^{(i)}] .
\label{SAMPLE}
\ee
As the number of measurements $N$ becomes large the quantity $\bar \Gamma$
will become a  Gaussian distribution about a mean value.  Its standard
deviation is    \cite{GPLTASI}
\bee
\sigma^2_\Gamma = {1 \over N}({1 \over N}\sum_{i=1}^N|\Gamma[\phi^{(i)}]|^2 -
\bar \Gamma^2). 
\label{STDEV}
\ee
The idea  of essentially all simulation algorithms
is that one constructs a new configuration
of field variables from an old one.  One begins with some simple field
configuration and monitors observables while the algorithm steps
along. After some number of steps, the value of observables will appear
to become independent of the starting configuration. At that point the
system is said to be ``in equilibrium'' and Eq. \ref{SAMPLE} can be
used to make measurements.

The simplest method for generating configurations is called the 
Metropolis  \cite{METROPOLIS}
algorithm. It works as follows:
From the old configuration $\{\phi\}$ with action $\beta S$, transform
the variables (in some reversible way) to a new trial configuration
$\{\phi\}'$ and compute the new action $\beta S'$. Then, if
$S' < S$ make the change and update all the variables; if not,
make the change with probability $\exp(-\beta(S'-S))$.

Why does it work? In equilibrium, the rate at which configurations $i$
turn into configurations $j$ is the same as the rate for the
back reaction $j \rightarrow i$. The rate of change is 
(number of configurations) $\times$ (probability of change). Assume
for the sake of the argument that $S_i < S_j$. Then the
rate $i \rightarrow j$ is $N_i P(i\rightarrow j)$ with
$P(i\rightarrow j)= \exp(-\beta(S_j-S_i)$ and  the
rate $j \rightarrow i$ is $N_j P(j\rightarrow i)$ with
$P(j\rightarrow i)= 1$. Thus $N_i/N_j = \exp(-\beta(S_i-S_j))$.

If you have any interest at all in the techniques I am describing, you should
write a little Monte Carlo program to simulate the two-dimensional
Ising model. Incidentally, the
 favorite modern method for pure gauge models is overrelaxation  \cite{OVERRELAX}.

One complication for QCD which spin models don't have is
fermions. The fermion path integral is not a number and a computer can't
simulate fermions directly.  However, one can formally integrate out the
fermion fields. For $n_f$ degenerate flavors of staggered fermions
\bee
Z = \int [dU][d\psi][d\bar\psi] \exp(-\beta S(U) - \sum_{i=1}^{n_f}
\bar \psi M \psi)
\ee
\bee
=\int [dU](\det M)^{n_f/2}\exp(-\beta S(U)) .
\ee
(One can make the determinant positive-definite by
writing it as $\det(M^\dagger M)^{n_f/4}$.)
The determinant introduces a nonlocal interaction among the $U$'s:
\bee
Z = \int [dU] \exp(-\beta S(U)
 - {n_f \over 4} {\rm Tr} \ln (M^\dagger M)
 ) .  
\ee

All large scale dynamical fermion simulations today generate
configurations using some variation of the microcanonical ensemble. That is,
they introduce momentum variables $P$ conjugate to the $U$'s and integrate
Hamilton's equations through a simulation time $t$
\bee
\dot U = i P U 
\ee
\bee
\dot P = -{{\partial S_{eff}}\over{\partial U}} . 
\label{ALLAT}
\ee
The integration is done numerically by introducing a timestep $\Delta t$.
The momenta are repeatedly refreshed by bringing them in contact with
 a heat bath and the method is thus called Refreshed or Hybrid Molecular
Dynamics  \cite{HMD}.

For special values of $n_f$ (multiples of 2 for Wilson fermions or
 of 4 for staggered fermions) the equations of motion can be derived
from a local Hamiltonian and in that case $\Delta t$ systematics
in the integration can be removed by an extra Metropolis accept/reject
step.  This method is called Hybrid Monte Carlo  \cite{HMC}.
 
The reason for the use of these small timestep algorithms is that for
any change in any of the $U$'s, $(M^\dagger M)^{-1}$ must be recomputed.
When Eq. \ref{ALLAT} is integrated all of the $U$'s in the lattice
are updated simultaneously, and only one matrix inversion is needed per
change of all the bosonic variables.
 
The major computational problem dynamical fermion simulations face is inverting
the fermion matrix $M$. It has eigenvalues with a very large range--
from $2\pi$ down to $m_q a$-- and in the physically interesting limit of
small $m_q$ the matrix becomes ill-conditioned.  At present it is necessary
to compute at unphysically heavy values of the quark mass and to extrapolate
to $m_q=0$.  The standard inversion technique today is one of the variants of
the  conjugate gradient algorithm \cite{FROMMER}.

\subsection{Taking The Continuum Limit, and Producing a Number in MeV}

When we define a theory on a lattice the lattice spacing 
 $a$ is an ultraviolet cutoff and all the coupling constants
in the action are  the  bare couplings 
defined with respect it.
  When we take $a$ to zero we must also specify how 
$g(a)$ 
behaves.  The proper continuum limit comes when we take $a$ to zero 
holding physical quantities fixed, not when we take $a$ to zero holding 
the couplings fixed.
 
On the lattice, if all quark masses are set to zero,
 the only dimensionful parameter is the lattice spacing, 
so all masses scale like $1/a$. Said differently, one computes
the dimensionless combination $am(a)$. One can determine the lattice spacing by
fixing one mass from experiment. Then all other dimensionful quantities
can be predicted.

Now imagine computing some masses at several values of the lattice spacing.
(Pick several values of the bare parameters at random and
calculate masses for each set of couplings.)
Our calculated mass ratios will depend on the lattice cutoff.
The typical behavior will look like
\bee
(a m_1 (a))/(a m_2 (a)) = m_1(0)/m_2(0) + O(m_1a) + O((m_1 a)^2) +\dots \label{SCALING}
\ee
The leading term does not depend on the value of the UV cutoff, while the
 other terms  do.
The goal of a lattice calculation (like the goal of almost any calculation
in quantum field theory) is to discover the value of some physical observable
as the UV cutoff is taken to be very large, so the physics is in the first term.
Everything else is an artifact of the calculation.
We say that a calculation ``scales'' if the $a-$dependent terms in
Eq. \ref{SCALING} are zero or small enough that one can extrapolate
to $a=0$, and generically refer to all the $a-$dependent terms
as ``scale violations.''

We can imagine expressing  each dimensionless combination $am(a)$
as some function of the bare coupling(s) $\{g(a)\}$, $am = f(\{g(a)\})$.
 As $a\rightarrow 0$ we must tune the set of couplings $\{g(a)\}$ so
\bee
 \lim_{a \rightarrow 0}
{1 \over a} f(\{g(a)\}) \rightarrow {\rm constant} . \label{2.35}
\ee
From the point of view of the lattice theory,
 we must tune $\{g\}$ so that correlation lengths $1/ma$ diverge.  
This will occur only at the locations of second (or higher) order phase 
transitions in the lattice theory.   

Recall that the $\beta$-function is defined by
 \bee
 \beta(g)=a{{dg(a)} \over da} ={dg(a) \over d {\rm ln}(1/\Lambda a)}. 
\label{2.36}
\ee
(There is actually one equation for each coupling constant in the set.
$\Lambda$ is a dimensional parameter introduced to make the argument of 
the logarithm dimensionless.)  At a critical point $\beta(g_c)$ = 0.  
Thus the continuum limit is the limit
\bee
\lim_{a \rightarrow 0}\{g(a)\} \rightarrow \{g_c\} .\label{2.37}
\ee
Continuum QCD is a theory with one dimensionless coupling
constant.
In QCD  the fixed point is $g_c = 0 $
so we must tune the coupling to vanish as $a$ goes to zero.

Pushing this a little further, the
two-loop $\beta$-function is prescription independent,
\bee
\beta(g) = -b_1 g^3 +b_2 g^5 ,\label{2.38}
\ee
and so if we think that the lattice theory is reproducing the continuum,
and if we think that the coupling constant is small enough that
the two-loop beta-function is correct,
we might want to observe perturbative scaling, or
``asymptotic scaling", $m/\Lambda$ fixed, or $a$ varying with $g$ as
\bee
a \Lambda = ({1 \over {g^2(a)}})^{b_2 /( 2 b_1^2 )}\exp(- {1 \over {b_1 
 g^2(a)}})
 . \label{2.39}
\ee

Asymptotic scaling is not scaling. Scaling means that
dimensionless ratios of physical observables do not depend on the
cutoff. Asymptotic scaling involves perturbation theory and the
definition of coupling constants. One can have one without the other.
(In fact, one can always define a coupling constant so that one
quantity shows asymptotic scaling.)

And this is not all.
There are actually two parts to  the problem of producing a number
 to compare with experiment.
One must first see scaling.
Then one needs to set the scale by
taking   some experimental number as input.
A complication that you may not have thought of
 is that the theory we simulate on the computer
is different from the real world. For example, 
a commonly used approximation is called the
``quenched approximation'': one neglects virtual quark loops, but includes
valence quarks in  the calculation.  The pion propagator is the propagator of
a $\bar q q$ pair, appropriately coupled, moving in a background of gluons.
This theory almost certainly does not have the same spectrum as QCD
with six flavors of dynamical quarks with their appropriate masses.
(In fact, an open question in the lattice community is, 
what is the accuracy of quenched approximation.)
Using one mass to set the scale from one of these
approximations to the real world might not give a prediction
for another mass which agrees with experiment.
We will see examples where this is important.

\subsection{Spectroscopy Calculations}

``In a valley something like a race took place. A little crowd watched
bunches of cars, each consisting of two `ups' and a `down' one, starting
in regular intervals and disappearing in about the same direction. `It
is the measurement of the proton mass,' commented Mr. Strange, `they
have done it for ages. A very dull job, I am glad I am not in the game.' ''
 \cite{SHURYAK}

Masses are computed in lattice  simulations from the
asymptotic behavior of Euclidean-time
 correlation functions.  A typical (diagonal) correlator can be written as
\bee
C(t) = \langle 0 | O(t) O(0) | 0\rangle  .  
\ee
Making the replacement
\bee
O(t)=e^{Ht}Oe^{-Ht} 
\ee
and inserting a complete set of energy eigenstates,  Eq. (3.1) \ becomes
\bee
C(t) = \sum_n |\langle 0 |  O|n\rangle |^2 e^{-E_nt}.  
\ee
At large separation the correlation function is approximately
\bee
C(t) \simeq  |\langle 0 |  O|1\rangle |^2 e^{-E_1t} 
\label{CORRFN}
\ee
where $E_1$ is the energy of the lightest state which the operator $O$
can create from the vacuum. 
If the operator does not couple to the vacuum, then
in the limit  of large $t$ one hopes to to find
the mass $E_1$
by measuring the leading exponential falloff of the correlation function,
and most lattice simulations begin with that measurement.
If the operator $O$ has poor overlap with the lightest state,
a reliable value for the mass can be extracted only at a large time $t$.
In some cases that state is the vacuum itself,
in which $E_1 = 0$.  
Then one looks for the next higher state--a signal which disappears
into the constant background.
This makes the actual calculation of the energy  more difficult.
 
This is the basic way hadronic masses are found in lattice gauge theory.  The
many calculations differ in important specific details of choosing the
 operators
$O(t)$.

\subsection{Recent Results}

 Today's supercomputer
QCD simulations range from $16^3 \times 32$ to
 $32^3 \times 100$
points and run from hundreds (quenched) to thousands (full QCD) of
hours on the fastest supercomputers in the world. 

Results are presented in four common ways. Often one sees a plot
of some bare parameter vs. another bare parameter. This is not
very useful if one wants to see continuum physics, but it is how
we always begin. Next, one can plot a dimensionless
ratio as a function of the lattice spacing. These plots
represent quantities like Eq. \ref{SCALING}. Both axes can show
 mass ratios.
Examples of such plots  are the so-called
Edinburgh plot ($m_N/m_\rho$ vs. $m_\pi/m_\rho$) and the Rome plot
 ($m_N/m_\rho$ vs. $(m_\pi/m_\rho)^2$).   These plots can answer 
continuum questions
(how does the nucleon mass change if  the quark mass is changed?)
or can be used to show (or hide) scaling violations.
Plots of one quantity in MeV vs. another quantity in MeV are typically
rather heavily processed after the data comes off the computer.

Let's look at some examples of spectroscopy, done in the ``standard way,''
with the plaquette gauge action and Wilson or staggered quarks.
I will restrict the discussion to quenched simulations because
only there are the statistical errors small enough to be interesting to
a non-lattice audience.
Most dynamical fermion simulations are unfortunately so noisy that
it is hard to subject them to detailed questioning.

\begin{figure}
\centerline{\ewxy{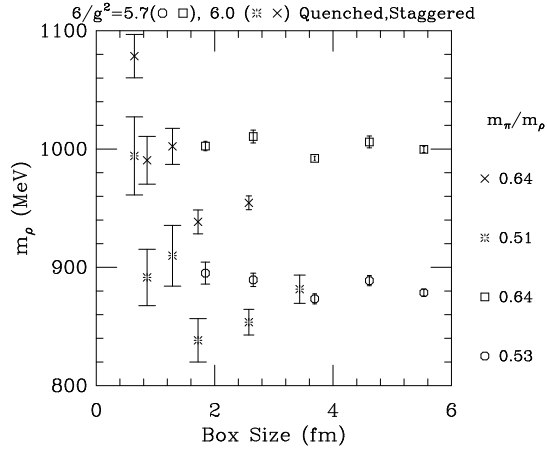}{80mm}
}
\caption{ Rho mass vs. box size.}
\label{fig:mrhovsnsa}
\end{figure}

Fig. \ref{fig:mrhovsnsa} shows a plot of the rho mass as a function
of the size of the simulation, for several values of the quark mass
(or $m_\pi/m_\rho$ ratio in the simulation) and lattice spacing
($\beta=6.0$ is $a \simeq 0.1$ fm and $\beta=5.7$ is about twice that)
 \cite{MILC}.
This picture shows that if the box has a diameter bigger than about 2 fm,
the rho mass is little affected, but if the box is made smaller, the rho is ``squeezed'' and its mass rises.

\begin{figure}
\centerline{\ewxy{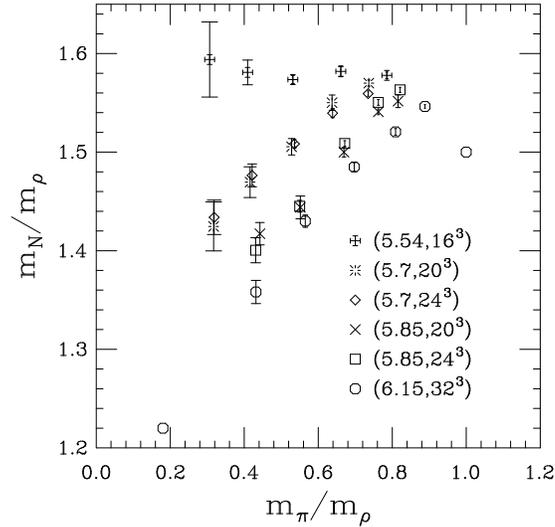}{80mm}
}
\caption{ An Edinburgh plot for staggered fermions, 
from the MILC collaboration.}
\label{fig:edinburgh}
\end{figure}

\begin{figure}
\centerline{\ewxy{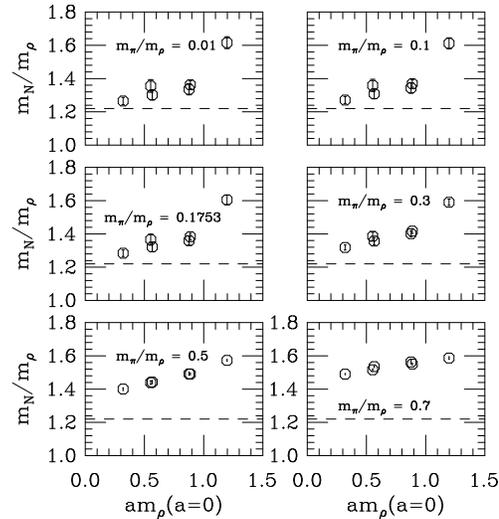}{80mm}
}
\caption{ ``Sections'' through the Edinburgh plot.}
\label{fig:fig3combo}
\end{figure}

Next we look at an Edinburgh plot, Fig. \ref{fig:edinburgh}  \cite{MILC}.
The different plotting symbols correspond to different bare couplings
or (equivalently) different lattice spacings.
This plot shows large scaling violations: mass ratios from different
lattice spacings do not lie on top of each other. We can expose the level of
scaling violations by taking ``sections'' through the plot and plot
$m_N/m_\rho$ at fixed values of the quark mass (fixed $m_\pi/m_\rho$), 
vs. lattice spacing, in
Fig. \ref{fig:fig3combo}.

\begin{figure}
\centerline{\ewxy{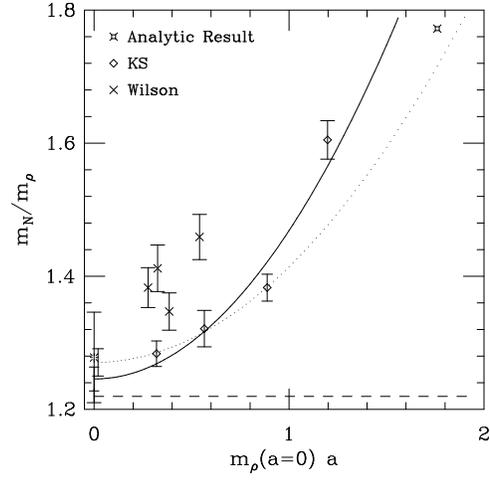}{80mm}
}
\caption{ Nucleon to rho mass ratio (at chiral limit) vs. lattice spacing
(in units of $1/m_\rho$).}
\label{fig:ratiovsmrhoa}
\end{figure}

Now for some examples of scaling tests in the chiral limit.
(Extrapolating to the chiral limit is a whole can of worms on its own,
but for now let's assume we can do it.)
Fig. \ref{fig:ratiovsmrhoa} shows the nucleon
to rho mass ratio (at chiral limit) vs. lattice spacing
(in units of $1/m_\rho$) for staggered  \cite{MILC} and Wilson   \cite{IBM}
fermions.
The ``analytic'' result is from strong coupling. The two curves
are quadratic extrapolations to zero lattice spacing using different
sets of points from the staggered data set. The burst is from a linear
extrapolation to the Wilson data. The reason I show this figure
is that one would like to know if the continuum limit of
quenched spectroscopy ``predicts'' the real-world $N/\rho$ mass ratio
of 1.22 or not. The answer (unfortunately) depends on how the reader
chooses to extrapolate.

\begin{figure}
\centerline{\ewxy{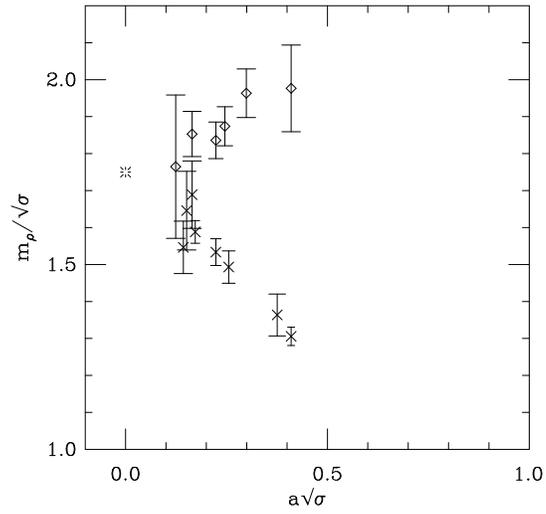}{80mm}
}
%
\caption{Scaling test for the rho mass in terms of the string tension,
with data points labeled as in Fig. 5.}
\label{fig:sommerfig}
\end{figure}

Another test  \cite{SOMMERPLOT}
 is the ratio of the rho mass to the square root of
the string tension, Fig. \ref{fig:sommerfig}.
 Here the diamonds are staggered data and the crosses
from the Wilson action. Scaling violations are large but the eye
extrapolates to something close to data (the burst).

\begin{figure}
\centerline{\ewxy{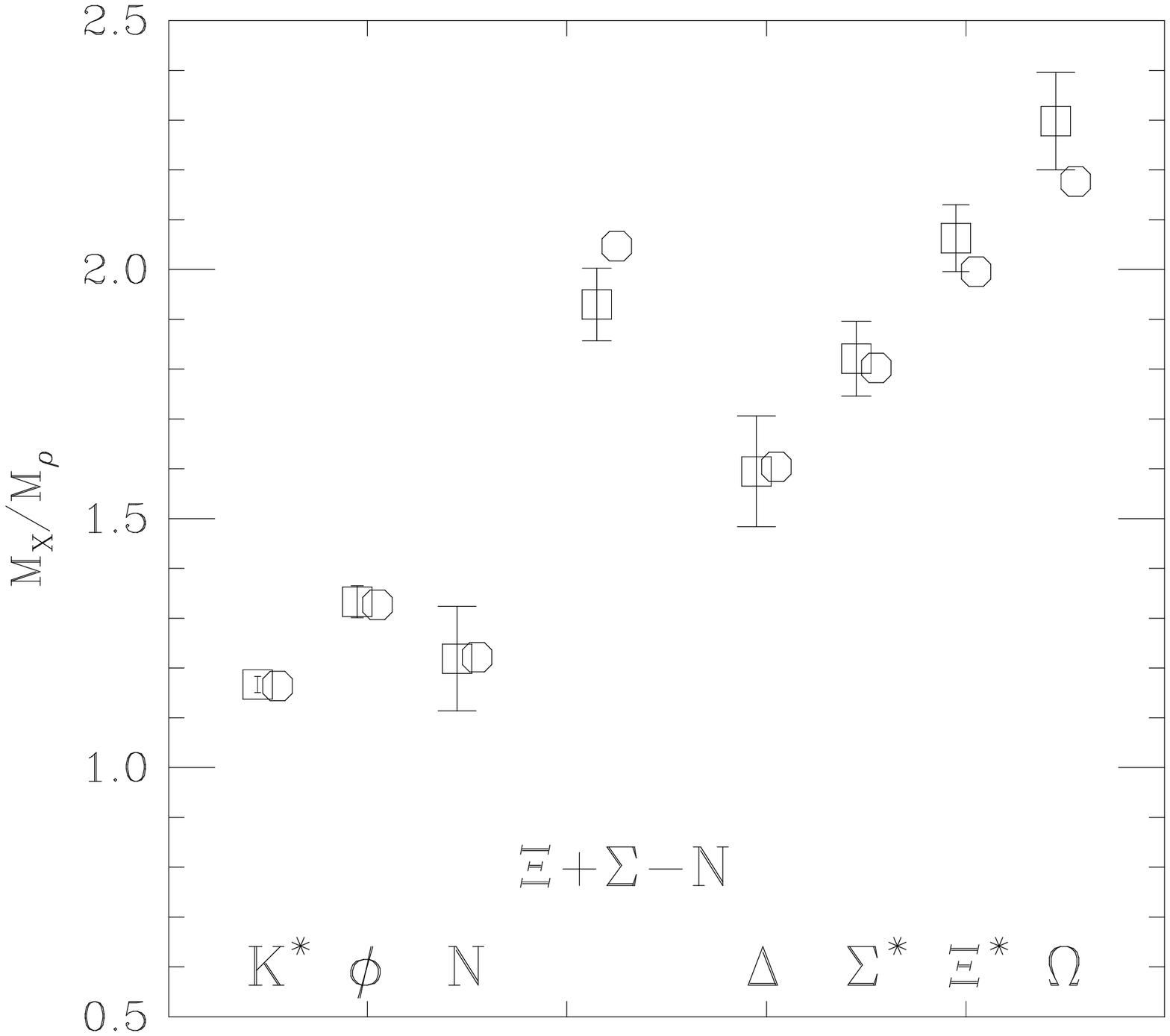}{80mm}
}
\caption{Quenched approximation mass ratios from Ref. 13.}
\label{fig:ratios}
\end{figure}

Finally, despite Mr. Strange, very few authors have attempted to
extrapolate to infinite volume, zero lattice spacing, and to physical quark
masses, including the strange quark. One group which did, 
Butler et al.  \cite{IBM}, produced Fig. \ref{fig:ratios}. The
squares are lattice data, the octagons are the real world. They look
quite similar within errors. Unfortunately, to produce this picture, they had
to build their own computer.

\section{Doing a Better Job--Maybe!}

The slow approach to scaling presents a practical problem for QCD
simulations, since it means that one needs to work at small lattice
spacing. This is expensive. The cost of a Monte Carlo simulation
in a box of physical size $L$ with lattice spacing $a$ and quark mass
$m_q$ scales roughly as
\bee
({L \over a})^4 ({1\over a})^{1-2}({1 \over m_q})^{2-3}
\label{COST}
\ee
where the 4 is just the number of sites, the 1-2 is the cost of
``critical slowing down''--the extent to which successive configurations
are correlated, and the 2-3 is the cost of inverting the fermion
propagator, plus critical slowing down from the nearly massless
pions. 
The problem  is that one needs a big computer to do anything.

However, all the simulations I described in the last lecture were
done with a particular choice of lattice action: the plaquette
gauge action, and either Wilson or staggered quarks. While those actions are
the simplest ones to program, they are just particular arbitrary choices of
bare actions. Can one invent a better lattice discretization, which
has smaller scaling violations?

People are trying many approaches. One could just write
down a slightly more complicated action, include some parameters which can be tuned, do a spectroscopy calculation, and see if there is any improvement
 as the parameters are varied. The problem with this method is
that it is like hunting for a needle in a multidimensional haystack--
there are so many possible terms to add. One needs an organizing
principle.

\subsection{Improvement based on naive dimensional analysis}

The simplest idea is to
use the naive canonical dimensionality of operators
 to guide us in our choice of improvement.
If we perform a naive Taylor expansion of a lattice operator like
the plaquette, we find that it can be written as
\beea
1 - {1 \over 3} {\rm Re \ Tr} U_{plaq} = &
                  r_0 {\rm Tr} F_{\mu\nu}^2  
+a^2 [ r_1 \sum_{\mu\nu}{\rm Tr} D_\mu F_{\mu\nu} D_\mu F_{\mu\nu} +
\nonumber \\
 &  r_2 \sum_{\mu\nu\sigma}{\rm Tr} D_\mu 
F_{\nu\sigma} D_\mu F_{\nu\sigma} + \nonumber \\
 &  r_3 \sum_{\mu\nu\sigma}{\rm Tr} D_\mu F_{\mu\sigma} D_\nu F_{\nu\sigma}]+
 \nonumber \\
 & +O(a^4) 
\eea
The expansion coefficients have a power series expansion in the
coupling, $r_j = A_j + g^2 B_j + \dots$
and the expectation value of any operator $T$ 
computed using the plaquette action will have an expansion
\bee
\langle T(a) \rangle = \langle T(0) \rangle + O(a) + O(g^2 a) + \dots
\ee

Other loops have a similar expansion, with different coefficients.
Now the idea is to take the lattice action to be
a minimal subset of loops and systematically remove the
 $a^n$ terms for physical observables order by order in $n$
 by taking the right linear combination of
loops in the action. 
\bee
S = \sum_j c_j O_j
\label{EXPA1}
\ee
with
\bee c_j = c_j^0 + g^2 c_j^1 + \dots
\label{EXPA2}
\ee 
This method was developed by Symanzik and 
co-workers  \cite{SYMANZIK,WEISZ,LWPURE} ten years ago.

\begin{figure}
\centerline{\ewxy{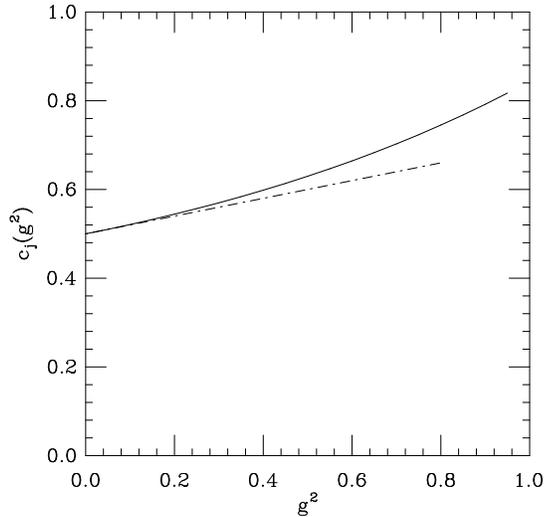}{80mm}
}
\caption{ The value of some parameter in a lattice action for
which physical observables have no $a^n$ errors. The dotted line
is the lowest order perturbative expectation.}
\label{fig:coupspace}
\end{figure}

To visualize this technique, look at Fig. \ref{fig:coupspace}.
We imagine parameterizing the coefficients of various terms in the
lattice action, which for a pure gauge theory could be a simple plaquette,
a $1 \times 2$ closed loop, the square of the $1 \times 2$ loop, and so on,
as some function of $g^2$.  ``Tree-level improvement'' involves specifying the
value of the j-th coefficient $c_j(g^2)$ at $g^2=0$. As we move away
 from $g^2=0$, the value of $c_j(g^2)$ for which observables calculated
using the lattice action have no errors through the specified exponent $n$
(no $a^n$ errors) will trace out a trajectory in coupling constant space.
For small $g^2$, the variation should be describable by perturbation
theory, Eq. \ref{EXPA2}, but when $g^2$ gets large, we would not
expect perturbation theory would be a good guide.

\begin{figure}
\centerline{\ewxy{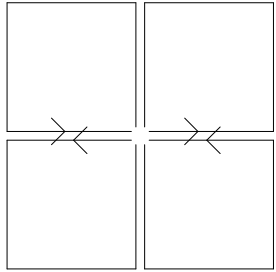}{80mm}
}
\caption{ The ``clover term''.}
\label{fig:clover}
\end{figure}

The most commonly used ``improved'' fermion action
is the
``Sheikholeslami-Wohlert''  \cite{SHWO} or ``clover'' action, an order $a^2$
improved Wilson action. The original
Wilson action has $O(a)$ errors in its vertices, $S_W = S_c + O(a)$.
This is corrected by making a field redefinition
\beea
\psi(x) \rightarrow \psi'(x) = & \psi(x) + {{ia}\over 4} \gamma_\mu D_\mu\psi
\\
\bar\psi(x) \rightarrow \bar\psi'(x) = & \bar \psi(x) + {{ia}\over 4} \gamma_\mu \bar\psi D_\mu
\eea
and the net result is an action with an extra lattice anomalous magnetic
moment term,
\bee
S_{SW} - {{iag}\over 4} \bar \psi (x)\sigma_{\mu\nu}F_{\mu\nu} \psi(x)
\ee
It is called the ``clover'' action because the lattice version
of $F_{\mu\nu}$ is the sum of paths shown in Fig. \ref{fig:clover}.

Studies performed at the time showed that this program
 did not improve scaling for the pure gauge theory
(in the sense that the cost of simulating the more complicated action 
was greater than the savings from using a larger lattice spacing.)
The whole program was re-awakened in the last few last years
 by Lepage and collaborators  \cite{PETERIMP} and variations of this program
give the most widely used ``improved'' lattice actions.

\subsection{Nonperturbative determination of coefficients}
\begin{figure}
\centerline{\ewxy{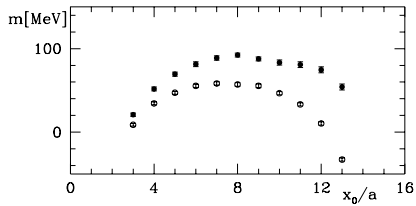}{80mm}
}
\caption{ Values of the quark mass as computed from the axial
and pseudoscalar currents, using the Wilson action. The open
and full symbols correspond to different boundary conditions on the
gauge fields.}
\label{fig:before}
\end{figure}

\begin{figure}
\centerline{\ewxy{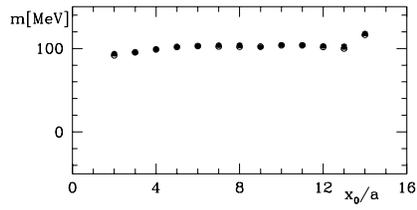}{80mm}
}
\caption{ Same as previous figure, but now with improved action
and operators.}
\label{fig:after}
\end{figure}

Although I am breaking chronological order, the simplest approach
to Symanzik improvement is the newest. The idea   \cite{NPIMP} is to force
the lattice to obey various desirable identities to some order in $a$,
by tuning parameters until the identities are satisfied by the simulations.
That is, we try to find the solid line in Fig, \ref{fig:coupspace}
by doing simulations.
Then use the action to
calculate other things and test to see if scaling
is improved. One example is the PCAC relation
\bee
\partial_\mu A_\mu^a = 2m_ P^a + O(a),
\ee
where the axial and pseudoscalar currents are just
\bee
A_\mu^a(x) = \bar \psi(x) \gamma_\mu\gamma_5 {1\over 2} \tau^a \psi(x)
\ee
and
\bee
P^a(x) = \bar \psi(x) \gamma_5 {1\over 2} \tau^a \psi(x)
\ee
($\tau^a$ is an isospin index.) The PCAC relation for the quark mass is
\bee
m \equiv  {1\over 2} {{\langle \partial_\mu A_\mu^a O^a \rangle} \over
{\langle P^a O^a \rangle}} + O(a).
\label{quarkmass}
\ee
Now the idea is to take some Symanzik-improved action, with
the improvement coefficients allowed to vary, 
and perform simulations in a little box with some particular choice of
boundary conditions for the fields. Parameters which can be 
tuned include the $c_{SW}$ in the clover term
 $i/4 c_{SW} a \sigma_{\mu\nu}F_{\mu\nu}$ and ones used for more complicated
expressions for the  currents
\bee
A_\mu = Z_A[(1+b_A a m_q)A_\mu^a + c_A a \partial_\mu P^a
\ee
\bee
P^a = Z_P(1+ b_P a m_q)P^a.
\ee
They are varied
until the quark mass, defined in Eq. \ref{quarkmass}, is independent
of location in the box, or of the boundary conditions.
Figs. \ref{fig:before} and \ref{fig:after} illustrate what can be done
with this tuning procedure.  It is still too soon for definitive tests
of scaling with this procedure.

\subsection{Improving perturbation theory}
\begin{figure}
\centerline{\ewxy{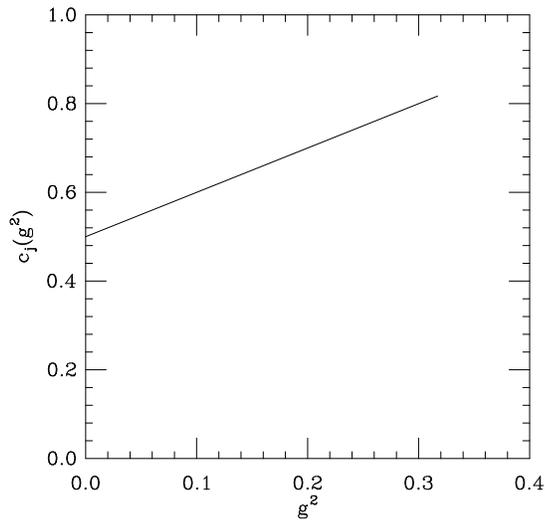}{80mm}
}
\caption{ Same as Fig. 8, but now with a redefined coupling constant $g^2_R$,
to make the ``improvement line'' linear in $g^2_R$.}
\label{fig:coupspace2}
\end{figure}

The older version of Symanzik improvement uses lattice perturbation
theory to compute the coefficients of the operators in the action.
The idea here is to find a new definition of $g^2$ for which the
solid line in Fig. \ref{fig:coupspace} is transformed into a straight line.
(Compare Fig. \ref{fig:coupspace2}.)

Let's make a digression into lattice perturbation theory \cite{MORNINGSTAR}.
It has three major uses.
First, we need to
relate lattice quantities (like matrix elements) to continuum ones:
$O^{cont}(\mu) = Z(\mu a, g(a))O^{latt}(a).$ 
This happens because the renormalization of an operator is slightly different
in the two schemes. In perturbation theory $Z$ has an expansion in powers of
$g^2$.
Second, we can use perturbation theory
to understand and check numerical calculations when the lattice 
couplings are very small.
Finally, one can use perturbative ideas to
 motivate nonperturbative improvement schemes \cite{PETERPAUL}.

Perturbation theory for lattice actions is just like any other
kind of perturbation theory (only much messier). One expands the
Lagrangian into a quadratic term and interaction terms
and constructs the propagator from the quadratic terms:
\beea
{\cal L} = & A_\mu(x) \rho_{\mu\nu}(x-y) A_\nu(y) + g A^3 + \dots \\
          =& {\cal L}_0 + {\cal L}_I .
\eea
For example, the gluon propagator in Feynman gauge for the Wilson
action is
\bee
D_{\mu\nu}(q) = {{g_{\mu\nu}}\over{ \sum_\mu(1-\cos(q_\mu a))}}.
\ee
To do perturbation theory for any system (not just the lattice)
one has to do three things:
one has to fix the renormalization scheme (RS) (define a coupling), 
specify the scale
at which the coupling is defined, and determine a numerical value for the 
coupling at that scale. All of these choices are arbitrary, and any perturbative
calculation is intrinsically ambiguous.

 Any object which has a perturbative expansion
can be written
\bee
O(Q) = c_0 + c_1(Q/\mu,RS) \alpha_s(\mu,RS) +c_2(Q/\mu,RS) \alpha_s(\mu,RS)^2
+\dots
\ee
In perturbative calculations we truncate the series after a fixed number
of terms and implicitly assume that's all there is. The  coefficients
$c_i(Q/\mu,RS)$ and the  coupling $\alpha_s(\mu,RS)$ depend on
the renormalization scheme  and choice of scale $\mu$.
 The guiding rule of perturbation theory \cite{MORNINGSTAR} is
``For a good choice of expansion the uncalculated higher order terms
should be small.''
 A bad choice has big coefficients.

There are many ways to define a coupling:
The most obvious is the bare coupling; as we will see shortly,
it is a poor expansion parameter. Another possibility is to define the
coupling from some physical observable. One popular choice is to use
the heavy quark potential at very high momentum transfer to define
\bee
V(q) \equiv 4 \pi C_f {{\alpha_V(q)}\over q^2}.
\ee
There are also several possibilities for picking a scale:
One can use the bare coupling, then $\mu=1/a$ the lattice spacing.
One can guess the scale or or play games just like in the continuum.
One game is the Lepage-Mackenzie $q^*$ prescription: find the ``typical''
momentum transfer $q^*$ for a process involving a loop graph by pulling
$\alpha_s(q)$ out of the loop integral and set
\bee
\alpha_s(q^*)\int d^4 q \xi(q) = \int d^4 q \xi(q) \alpha_s(q).
\label{BLME}
\ee
To find $q^*$, write $\alpha_s(q) = \alpha_s(\mu) +
b \ln(q^2/\mu^2)\alpha_s(\mu)^2 +\dots$,
and similarly for $\alpha_s(q^*)$, insert these expressions
into Eq. \ref{BLME} and compare the
$\alpha_s(\mu)^2$ terms, to get
\bee
\ln(q^*) = \int d^4 q \ln(q) \xi / \int d^4 q \xi .
\label{IMMP}
\ee
This is the lattice analog of the Brodsky-Lepage-Mackenzie  \cite{BLM}
prescription in continuum PT.

\begin{figure}
\centerline{\ewxy{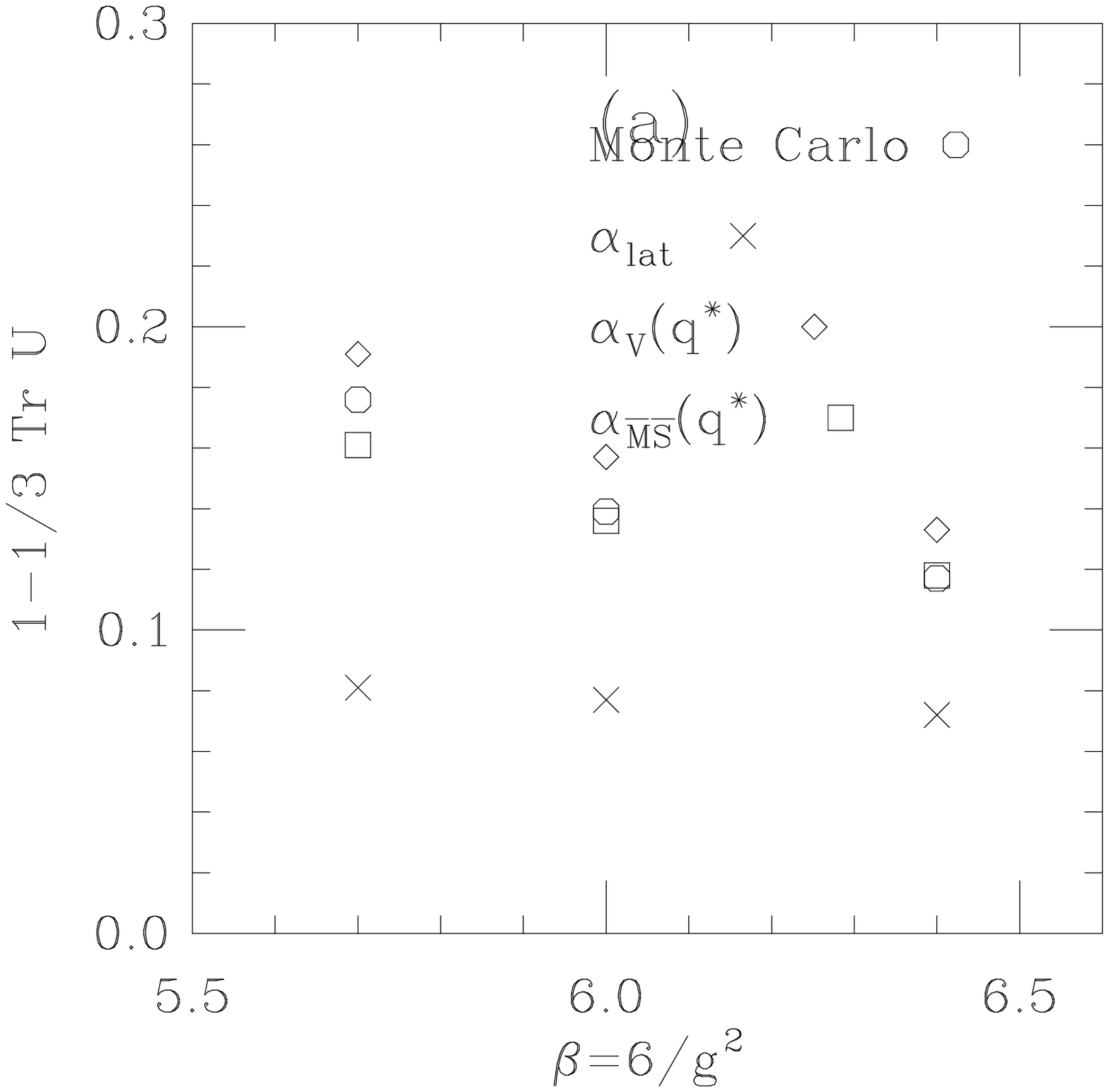}{80mm}
\ewxy{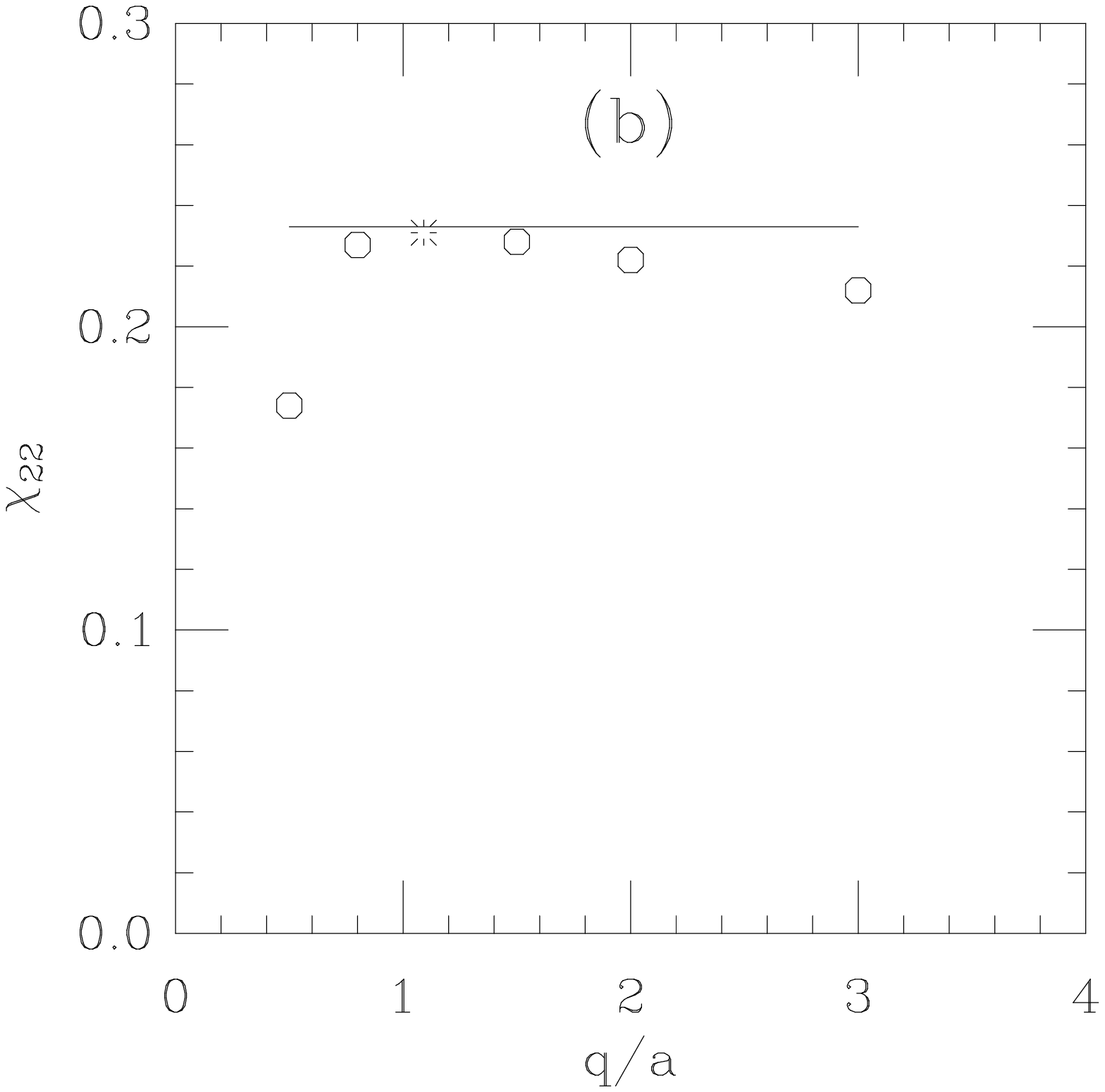}{80mm}}
\vspace{0.5cm}
\centerline{\ewxy{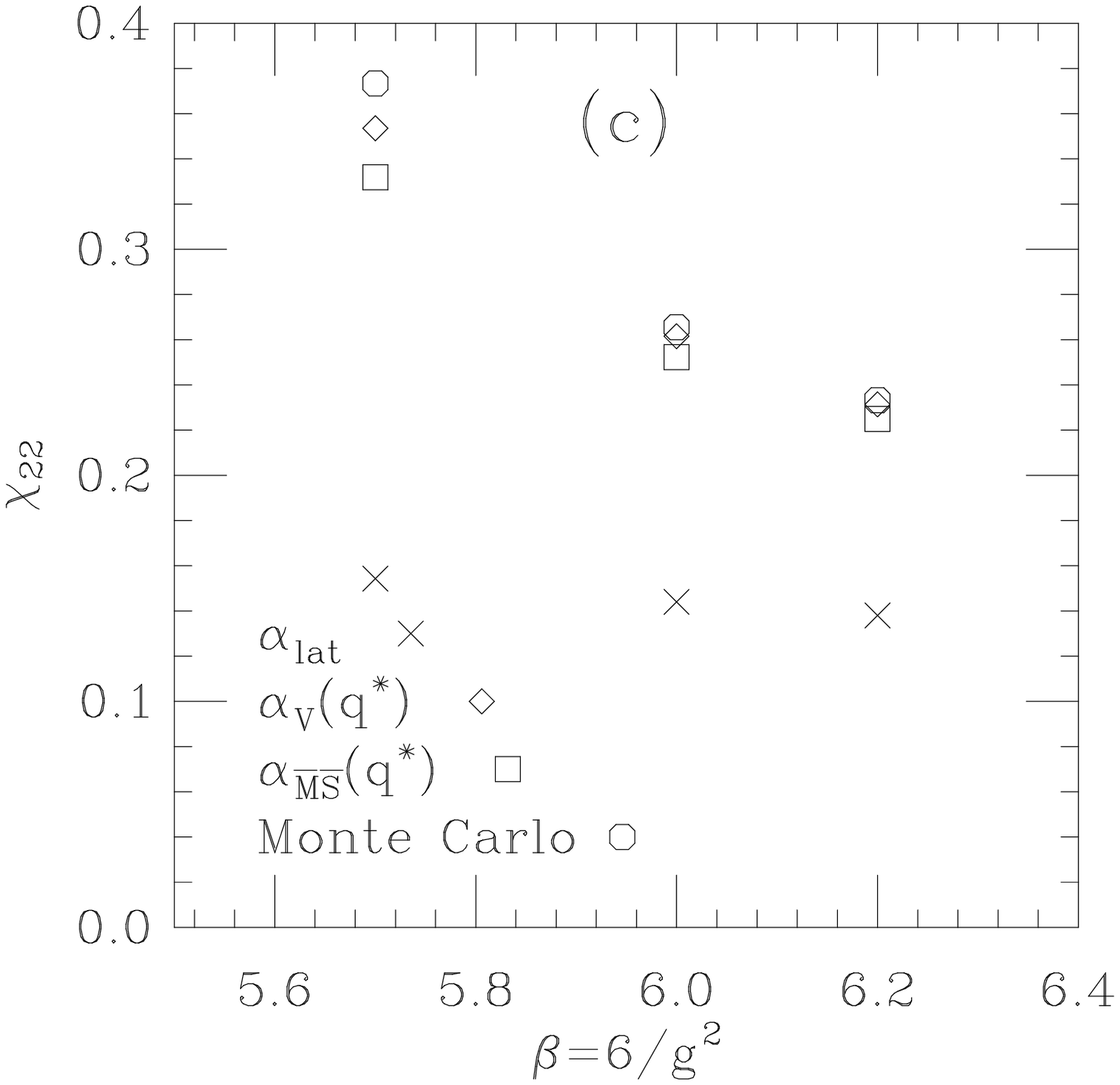}{80mm}
}
\caption{ Examples of ``improved perturbation theory''.}
\label{fig:improve}
\end{figure}

Finally one must determine the coupling:
If one uses the bare lattice coupling it is already known. Otherwise,
one can compute it in terms of the bare coupling:
\bee
\alpha_{\overline{MS}}(s/a) = \alpha_0 + (5.88-1.75 {\rm ln} s) \alpha_0^2
+ (43.41 - 21.89 {\rm ln} s + 3.06 {\rm ln}^2 s) \alpha_0^3 + \dots
\ee
Or one can determine it from something one measures on the lattice,
which has a perturbative expansion. For example
\bee
-{\rm ln}\langle {1\over 3} {\rm Tr}U_{plaq} \rangle=
{4\pi \over 3}\alpha_P(3.41/a)(1-1.185\alpha_P)
\label{PLAQALPHA}
\ee
(to this order, $\alpha_P=\alpha_V$).
Does ``improved perturbation theory'' actually improve
 perturbative calculations?
In many cases, yes: some examples are shown in Fig. \ref{fig:improve}
from  \cite{PETERPAUL}:
On the upper left we see a calculation of the average link in Landau
gauge, from simulations (octagons) and then from lowest-order
perturbative calculations using the bare coupling (crosses)
and $\alpha_V$ and $\alpha_{\overline{MS}}$ (diamonds and squares).
In the upper right panel we see how the lattice prediction of
 an observable involving the 2 by 2 Wilson loop depends on the
choice of momentum $q^*/a$ (at $\beta=6.2$, a rather weak value of the
bare coupling)
in the running coupling constant. The
burst is the value of the prescription of Eq. \ref{IMMP}.
In the lower panel are perturbative predictions the same observables
as a function of lattice coupling. These pictures illustrate
that perturbation theory in terms of the bare coupling does not work well,
but that using other definitions for
couplings, one can get much better agreement with
the lattice ``data''.

\begin{figure}
\centerline{\ewxy{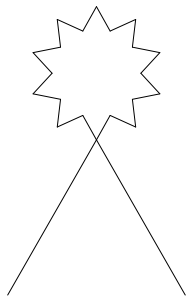}{80mm}
}
\caption{ The ``tadpole diagram''.}
\label{fig:tadpole}
\end{figure}

Straight perturbative expansions by themselves for the commonly-used lattice
actions  are typically not very convergent. The culprit is
the presence of $U_\mu$'s in the action. One might think that for weak
coupling, one could expand
\bee
\bar\psi U \psi = \bar \psi [ 1 + iga A + \dots ] \psi
\ee
and ignore the $\dots$, 
but the higher order term $\bar \psi {1\over 2}g^2a^2 A^2   \psi$
generates the ``tadpole graph'' of Fig. \ref{fig:tadpole}.
The UV divergence in the gluon loop $\simeq 1/a^2$ cancels the
$a^2$ in the vertex. The same thing happens at higher order,
and the tadpoles add up to an effective $a^0\sum c_ng^{2n}$ contribution.
Parisi  \cite{PARISI} and later
Lepage and Mackenzie  \cite{PETERPAUL} suggested a heuristic way
to deal with this problem: replace
$U_\mu \rightarrow u_0(1+igaA)$ where $u_0$, the ``mean field term'' or
``tadpole improvement term'' is introduced phenomenologically to sum the loops.
Then one rewrites the action as
\bee
S = {1 \over {g^2 u_0^4}} \sum {\rm Tr}U \leftrightarrow {1\over{g_{lat}^2}}
\sum {\rm Tr U}
\ee
where $g^2 \equiv g_{lat}^2/u_0^4$ is the new expansion parameter.
Is $u_0^4 \equiv \langle {\rm Tr}U_{plaq}/3\rangle$? This choice is often
used; it is by no means unique.

A ``standard action'' (for this year, anyway) is the
``tadpole-improved L\"uscher-Weisz  \cite{LWPURE}
 action,'' composed of a 1 by 1, 
1 by 2, and ``twisted'' loop (+x,+y,+z,-x,-y,-z),
\bee
\beta S = -\beta [ {\rm Tr}(1\times 1) - {1 \over {20 u_0^2}}
(1+0.48 \alpha_s){\rm Tr}(1 \times 2)
- {1 \over u_0^2} 0.33 \alpha_s {\rm Tr}U_{tw} ]
\ee
with $u_0 \equiv  \langle {\rm Tr}U_{plaq}/3\rangle^{1/4}$
and $3.068 \alpha_s \equiv -\ln\langle {\rm Tr}U_{plaq}/3\rangle$
determined self-consistently in the simulation.

\subsection{Fixed Point Actions}
Let's recall the question we were trying to answer in the previous
sections: Can one find a trajectory in coupling constant space, along
which the physics has no corrections to some desired order in
$a^n$ or $g^m a^n$? Let's take the question one step further:
Is there a trajectory in coupling constant space in which there
are {\it no} corrections at all, for any $n$ or $m$?

To approach the answer, let's think about the connection
between scaling and the properties of some arbitrary bare action,
which we assume is defined with some UV cutoff $a$ (which does not
have to be a lattice cutoff, in principle). The action is
characterized by an infinite number of coupling constants, $\{c\}$.
(Many of them could be set to zero.)
When the $c$'s take on almost any arbitrary values, the typical scale for all
physics will be the order of the cutoff: $m \simeq 1/a$, correlation
length $\xi \simeq a$. There will be strong cutoff effects.

The best way to think about scaling is through the renormalization
group  \cite{WK}. Take the action with cutoff $a$ and integrate out degrees
of freedom to construct a new effective action with a new cutoff
$a' > a$ and fewer degrees of freedom. The physics at distance scales
$r > a$ is unaffected by the coarse-graining (assuming it is done
exactly.) We can think of the effective actions as being similar to the
original action, but with altered couplings. We can repeat this
coarse-graining and generate new actions with new cutoffs.
As we do, the coupling constants ``flow:''
\bee
S(a,c_j) \rightarrow S(a',c_j') \rightarrow S(a'',c_j'')\rightarrow \dots
\ee
If under repeated blockings the system flows to a fixed point
\bee
S(a_n,c^n_j) \rightarrow S(a_{n+1},c_j^{n+1}=c_j^n)
\ee
then observables are independent of the cutoff $a$ and in particular
the correlation length  $\xi$ must either be zero or infinite.

This can only happen if the original $c$'s belong to a particular
 restricted set,
called the ``critical surface.'' It is easy to see that physics
on the critical surface is universal: at long distances the effective
 theory is the action at the fixed point, to which all the couplings 
have flowed,
regardless of their original bare values. In particular, physics at the
fixed point is independent of the underlying lattice structure.

But $\xi = \infty$ is not $\xi$ large. Imagine tuning bare parameters 
close to the critical surface, but not on it. The system will 
flow towards the fixed point, then away from it.  The flow lines in coupling
constant space will asymptotically approach a particular trajectory,
called the renormalized trajectory (RT), which connects (at $\xi = \infty$)
with the fixed point. Along the renormalized trajectory, $\xi$ is finite.
However, because it is connected to the fixed point, it shares 
the scaling properties of the fixed point--in particular, the 
complete absence of cutoff effects in the spectrum and in Green's functions.
(To see this remarkable result, imagine doing  QCD
spectrum calculations with the original bare action with a cutoff
equal to the Planck mass and then coarse graining.
Now exchange the order of the two procedures.
 If this can be done without making any approximations the
answer should be the same.)

\begin{figure}
\centerline{\ewxy{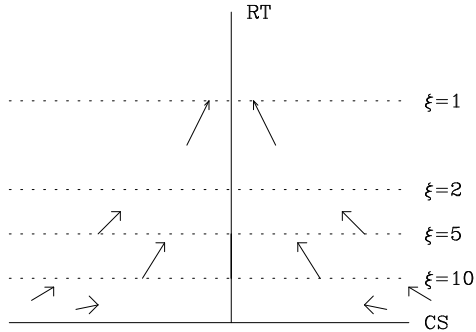}{80mm}
}
\caption{ A schematic picture of renormalization group flows along a one-dimensional critical surface, with the associated renormalized
trajectory, and superimposed contours of constant correlation length.}
\label{fig:rt}
\end{figure}

A Colorado analogy is useful for visualizing the
critical surface and renormalized trajectory: 
think of the critical surface as the top
of a high mountain ridge. The fixed point is a saddle point on the ridge.
A stone released on the ridge will roll to the saddle and come to rest.
 If it is not released exactly on the ridge, it will roll near
to the saddle, then go down the gully leading away from it.
For a cartoon, see Fig. \ref{fig:rt}.

So the ultimate goal of ``improvement programs'' is to find a true
perfect action, without cutoff effects, along the renormalized trajectory
of some renormalization group transformation. At present, finding
an RT has not been done in a convincing way for any renormalization
group transformation. However, an
 action at the fixed point might also be an improved action, and
fixed point actions really can be constructed and used.

 In lattice
language, a bare action for QCD is  described by one overall factor of 
$\beta=2N/g^2$
and arbitrary weights of various closed loops,
\bee
\beta S = {{2N} \over g^2} \sum_j c_j O_j.
\ee
Asymptotic freedom is equivalent to the statement that
the critical surface of any renormalization group transformation
is at $g^2=0$. The location of a fixed point
involves some relation among the $c_j$'s.

A  direct attack on the renormalized trajectory begins by
finding a fixed point action. Imagine having a set of
field variables $\{\phi\}$ defined with a cutoff $a$. Introduce
some coarse-grained variables $\{\Phi\}$ defined with respect to
a new cutoff $a'$, and integrate out the fine-grained variables to
produce a new action
\bee
\rme^{-\beta S'(\Phi)} = \int d\phi  \rme^{-\beta(T(\Phi,\phi)+S(\phi))}
\label{RGE}
\ee
where $\beta(T(\Phi,\phi)$ is the blocking kernel which
functionally relates the coarse and fine variables. Integrating
Eq. \ref{RGE} is usually horribly complicated. However, P. Hasenfratz
and F. Niedermayer   \cite{HN} noticed an amazing simplification for
asymptotically free theories: Their critical surface is at $\beta=\infty$
and in that limit Eq. \ref{RGE} becomes a steepest-descent relation
\bee
S'(\Phi) = \min_{\phi}((T(\Phi,\phi)+S(\phi))
\ee
which can be used to find the fixed point action
\bee
S_{FP}(\Phi) = \min_{\phi}((T(\Phi,\phi)+S_{FP}(\phi)).
\label{RGFP}
\ee
The program has been successfully carried out for $d=2$ 
sigma models  \cite{HN} and for four-dimensional pure gauge theories
 \cite{PAPER123}.  These actions have two noteworthy properties:
First, not only are they classically perfect actions (they have no
$a^n$ scaling violations for any $n$), but they are also one-loop
quantum perfect: that is, as one moves out the renormalized trajectory,
\bee
{1\over g^2} S_{RT}(g^2) = {1 \over g^2}(S_{FP} + O(g^4) ).
\label{QPF}
\ee
Physically this happens because the original action has no irrelevant
operators, and they are only generated through loop graphs.
Thus these actions are an extreme realization of the Symanzik
program. Second, because these actions are at the fixed point, they
have scale invariant classical solutions. This fact can be used
to define a topological charge operator on the lattice in
a way which is consistent with the lattice action  \cite{INSTANTON12}.

These actions are ``engineered'' in the following way: one picks
a favorite blocking kernel, which has some free parameters, and
solves Eq. \ref{RGFP}, usually approximately at first.
Then one tunes the parameters in the kernel to optimize the 
action for locality, and perhaps refines the solution.
Now the action is used in simulations at finite correlation
length (i.e. do simulations with a Boltzman factor
$\exp(-\beta S_{FP})$. Because of Eq. \ref{QPF}, one believes
that the FP action will be a good approximation to the
perfect action on the RT; of course, only a numerical test can tell.
As we will see in the next section, these actions perform very well.
At this point in time no nonperturbative
FP action which includes fermions has been tested, but most of
the formalism is there  \cite{WIESE}.

\subsection{Examples of ``Improved'' Spectroscopy}
I would like to show some examples
of the various versions of ``improvement'', 
and remind you of the pictures at the
end of the last chapter to contrast results from  standard actions.

\begin{figure}
\centerline{\ewxy{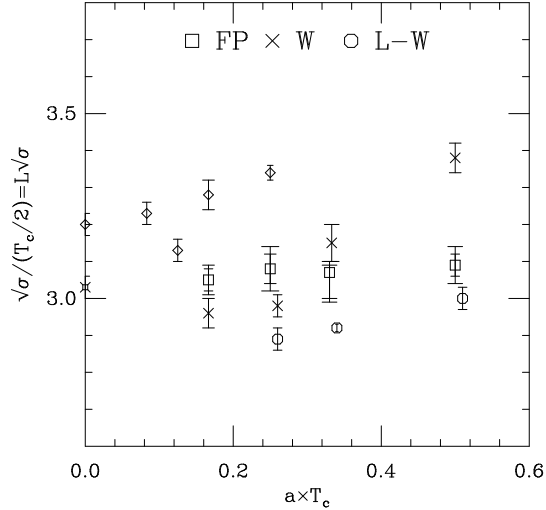}{80mm}
}
\caption{ The (square root of) the string tension in lattices of
constant physical size $L=2/T_c$,
but different lattice spacings (in units of $1/T_c$). 
}
\label{fig:aspect2b}
\end{figure}

Fig. \ref{fig:aspect2b} shows a plot of the string tension measured
in systems of constant physical size (measured in units of $1/T_c$,
the critical temperature for deconfinement), for SU(3) pure gauge theory.
In the quenched approximation, with $\sqrt{\sigma}\simeq 440$ MeV,
$T_c= 275$ MeV and $1/T_c=0.7$ fm.
Simulations with the standard Wilson action are crosses, while the squares show
FP action results  \cite{PAPER123}  and the octagons from the
tadpole-improved L\"uscher-Weisz action  \cite{BLISS}. 
The figure  illustrates that it is hard
to quantify improvement. There are no measurements with the
Wilson action at small lattice spacing of of precisely
the same observables that the ``improvement people'' measured. The best
one can do is to take similar measurements (the diamonds) and attempt
to compute the $a=0$ prediction for the observable we measured (the fancy
cross at $a=0$). This attempt lies on a straight line with the FP
action data, hinting strongly that the FP action is indeed scaling.
The FP action seems to have gained about
a factor of three to four in lattice spacing, or a gain of 
$(3-4)^6$ compared to the 
plaquette action, according to Eq. \ref{COST},
at a cost of a factor of 7 per site because it is more
complicated to program.
The tadpole-improved L\"uscher-Weisz action data lie lower than the
FP action data and do not scale as well.  
As $a\rightarrow 0$ the two actions should yield the same result;
that is just universality at work. However, there is no guarantee that
the approach to the continuum is monotonic.

\begin{figure}
\centerline{\ewxy{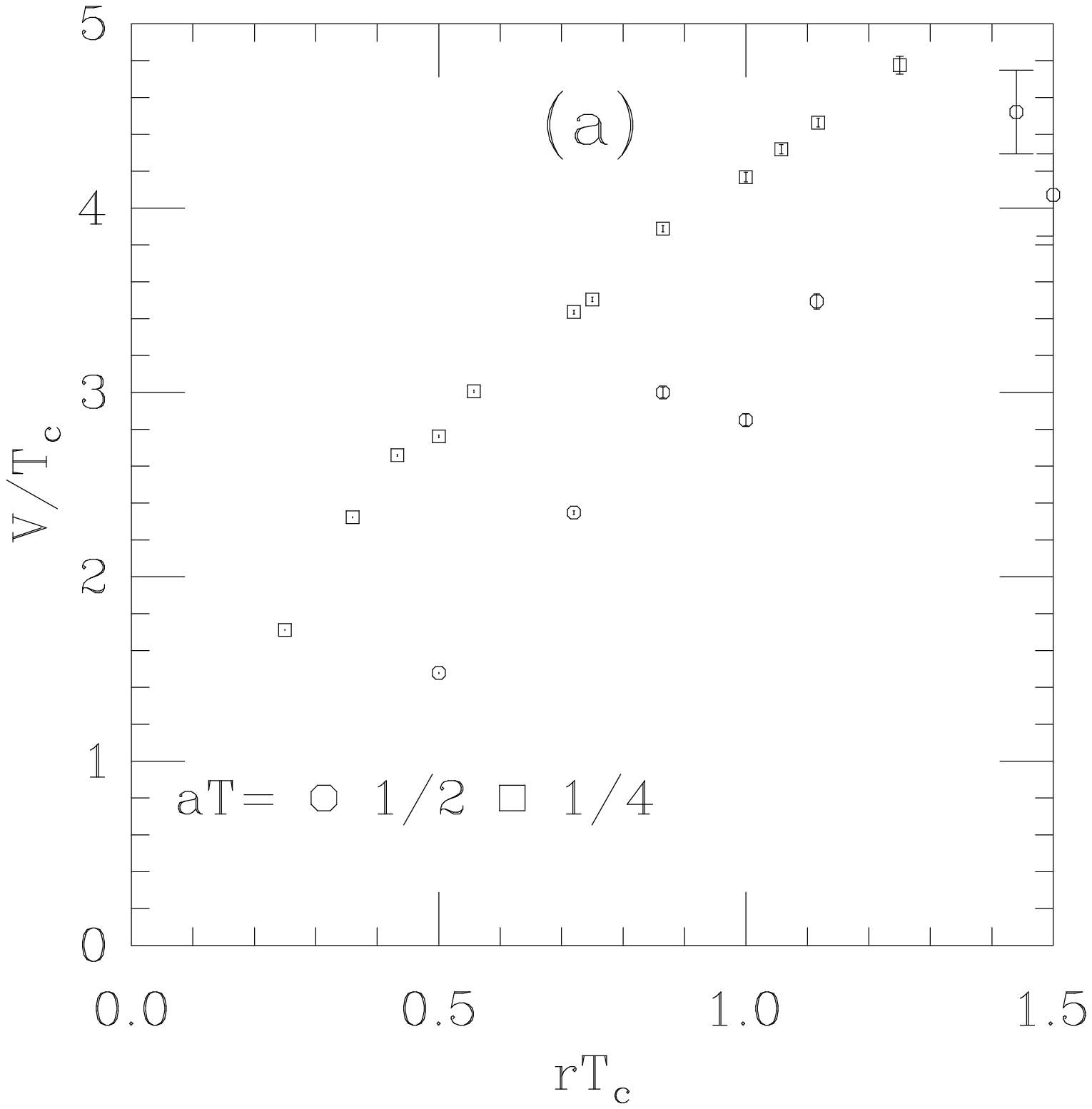}{80mm}
\ewxy{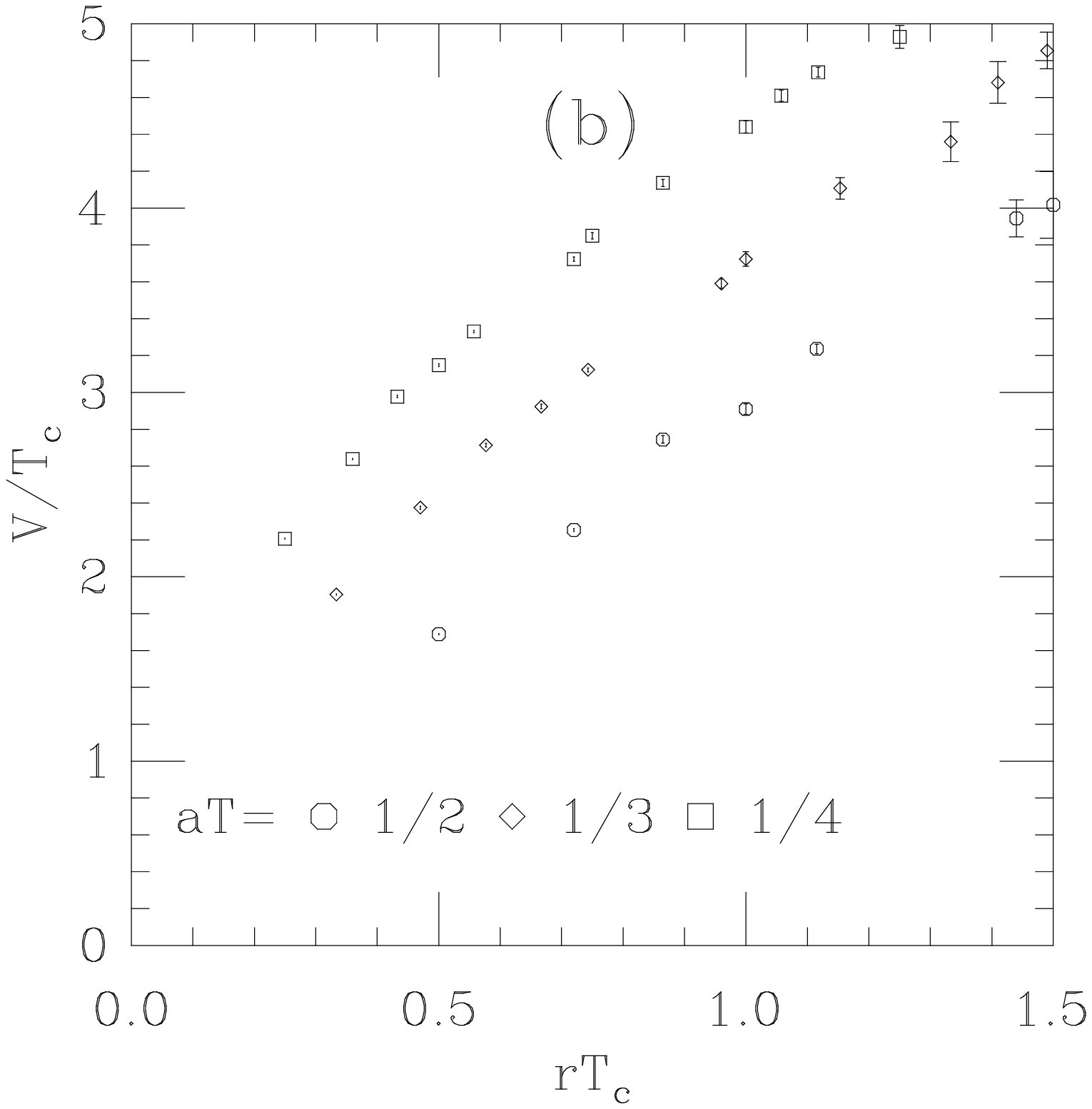}{80mm}}
\caption{ The heavy quark potential in SU(2) pure  gauge theory
measured in units of $T_c$. (a) Wilson action (b) an FP action.}
\label{fig:vr}
\end{figure}

Fig. \ref{fig:vr} shows the heavy quark-antiquark potential in SU(2)
gauge theory, where $V(r)$ and $r$ are measured in the appropriate units
of $T_c$, the critical temperature for deconfinement. The Wilson action
is on the left and a FP action is on the right. The vertical displacements
of the potentials are just there to separate them.
Notice the large violations of rotational symmetry in the Wilson
action data when the lattice spacing is $a= 1/2T_c$ which are
considerably improved in the FP action results.

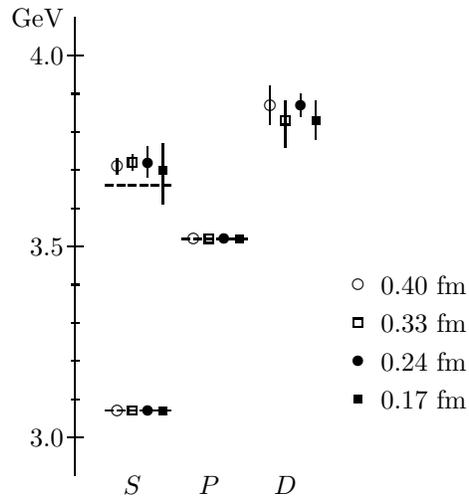
\begin{figure}
\begin{center}
\setlength{\unitlength}{.02in}
\begin{picture}(120,130)(0,280)
\put(89,340){\circle{3}}\put(95,340){\makebox(0,0)[l]{$0.40$~fm}}
\put(88,329){\framebox(2,2){\mbox{}}}
             \put(95,330){\makebox(0,0)[l]{$0.33$~fm}}
\put(89,320){{\circle*{3}}}\put(95,320){\makebox(0,0)[l]{$0.24$~fm}}
\put(88,310){\rule[-\unitlength]{2\unitlength}{2\unitlength}}
         \put(95,310){\makebox(0,0)[l]{$0.17$~fm}}

\put(15,290){\line(0,1){120}}
\multiput(13,300)(0,50){3}{\line(1,0){4}}
\multiput(14,310)(0,10){9}{\line(1,0){2}}
\put(12,300){\makebox(0,0)[r]{3.0}}
\put(12,350){\makebox(0,0)[r]{3.5}}
\put(12,400){\makebox(0,0)[r]{4.0}}
\put(12,410){\makebox(0,0)[r]{GeV}}

\put(30,290){\makebox(0,0)[t]{$S$}}

\multiput(23,307)(3,0){6}{\line(1,0){2}}
\put(26,307){\circle{3}}
\put(29,306){\framebox(2,2){\mbox{}}}
\put(34,307){\circle*{3}}
\put(37,306){\rule{2\unitlength}{2\unitlength}}

\multiput(23,366)(3,0){6}{\line(1,0){2}}
\put(26,371){\circle{3}}
\put(26,369){\line(0,1){4}}
\put(29,371){\framebox(2,2){\mbox{}}}
\put(30,370){\line(0,1){4}}
\put(34,372){\circle*{3}}
\put(34,368){\line(0,1){8}}
\put(37,369){\rule{2\unitlength}{2\unitlength}}
\put(38,361){\line(0,1){16}}

\put(50,290){\makebox(0,0)[t]{$P$}}

\multiput(43,352)(3,0){6}{\line(1,0){2}}
\put(46,352){\circle{3}}
\put(49,351){\framebox(2,2){}}
\put(54,352){\circle*{3}}
\put(54,351){\line(0,1){2}}
\put(57,351){\rule{2\unitlength}{2\unitlength}}
\put(58,351.5){\line(0,1){1}}

\put(70,290){\makebox(0,0)[t]{$D$}}

\put(66,387){\circle{3}}
\put(66,382){\line(0,1){10}}
\put(69,382){\framebox(2,2){}}
\put(70,376){\line(0,1){12}}
\put(74,387){\circle*{3}}
\put(74,384){\line(0,1){6}}
\put(77,382){\rule{2\unitlength}{2\unitlength}}
\put(78,378){\line(0,1){10}}
\end{picture}


\end{center}
\caption{$S$, $P$, and $D$ states of charmonium computed on lattices with:
$a=0.40$~fm (improved action, $\beta_{plaq}=6.8$);
$a=0.33$~fm (improved action, $\beta_{plaq}=7.1$);
$a=0.24$~fm (improved action, $\beta_{plaq}=7.4$); and
$a=0.17$~fm (Wilson action, $\beta=5.7$, from~[43]),
from Ref. 19. The dashed lines
indicate the true masses.}
\label{fig:spect}
\end{figure}

Next we consider nonrelativistic QCD. A comparison of the quenched 
charmonium spectrum from Ref. \cite{PETERIMP} using
data from Ref. \cite{REF6} is shown in
Fig. \ref{fig:spect}. When the tadpole-improved L-W action is used to generate
gauge configurations, the scaling window is pushed out to $a\simeq 0.4$
fm for these observables.

\begin{figure}
\centerline{\ewxy{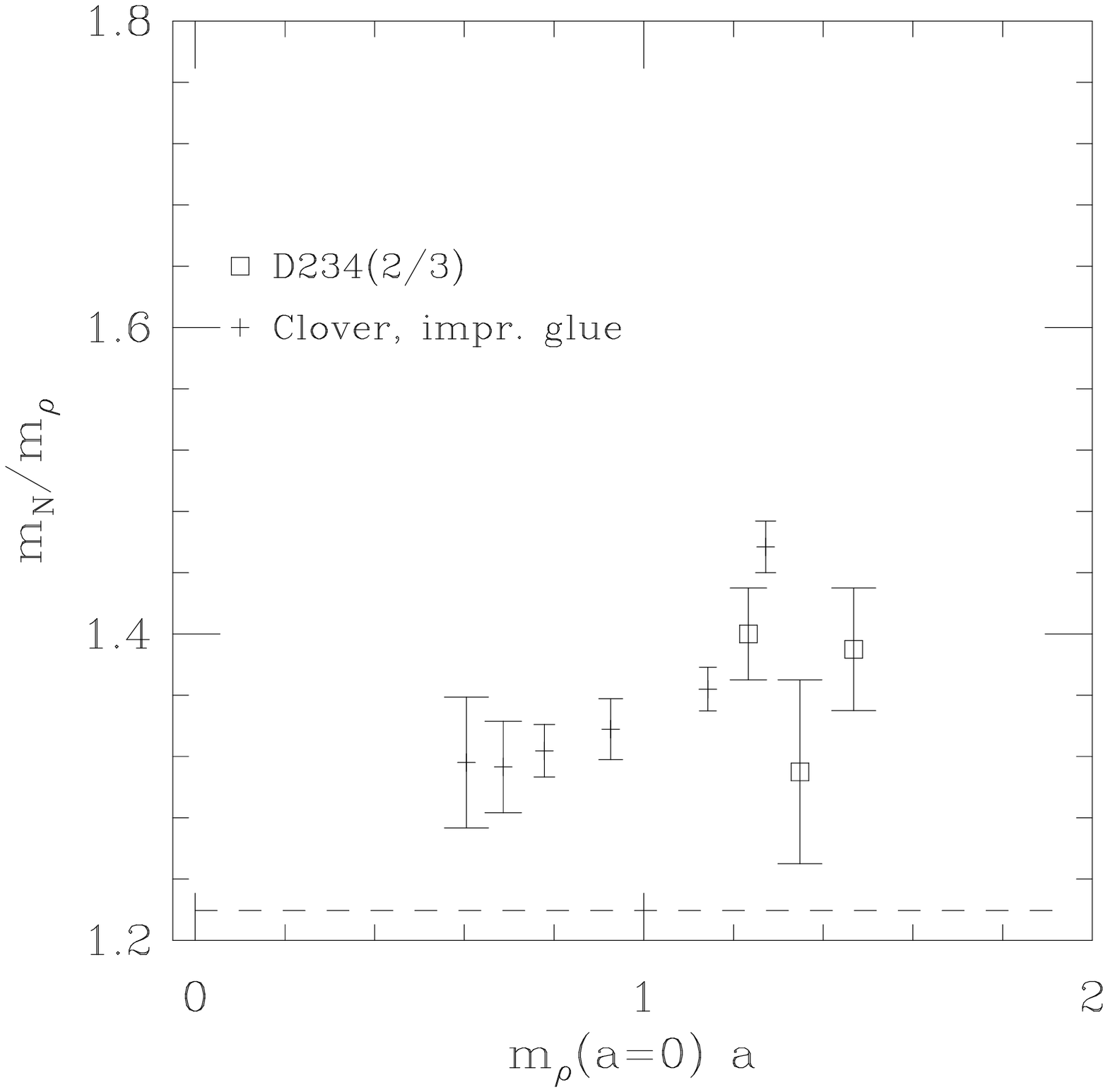}{80mm}
}
\caption{ Nucleon to rho mass ratio (at chiral limit) vs. lattice spacing
(in units of $1/m_\rho$).}
\label{fig:ratiovsmrhoa96}
\end{figure}

\begin{figure}
\centerline{\ewxy{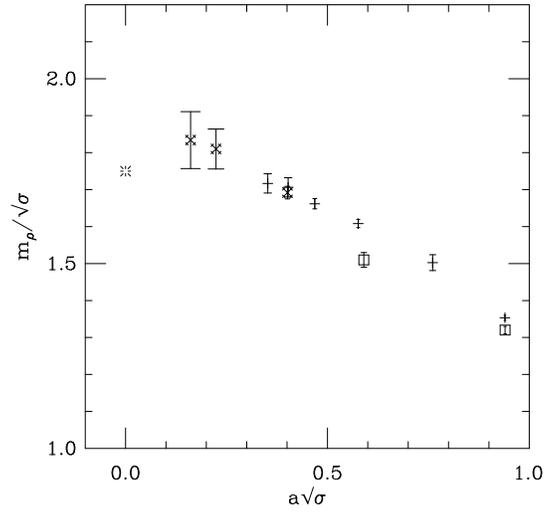}{80mm}
}
%
\caption{Rho mass scaling test with respect to the string tension. }
\label{fig:sommerfigim}
\end{figure}

\begin{figure}
\centerline{\ewxy{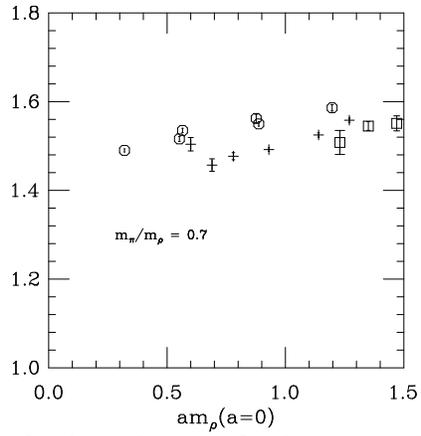}{80mm}
}
\caption{$m_N/ m_\rho$ vs $a m_\rho$ at fixed quark mass
(fixed $m_\pi/m_\rho$).  Interpolations of the S-W and D234(2/3)
data were done by me. }
\label{fig:ratioimp0.7}
\end{figure}

\begin{figure}
\centerline{\ewxy{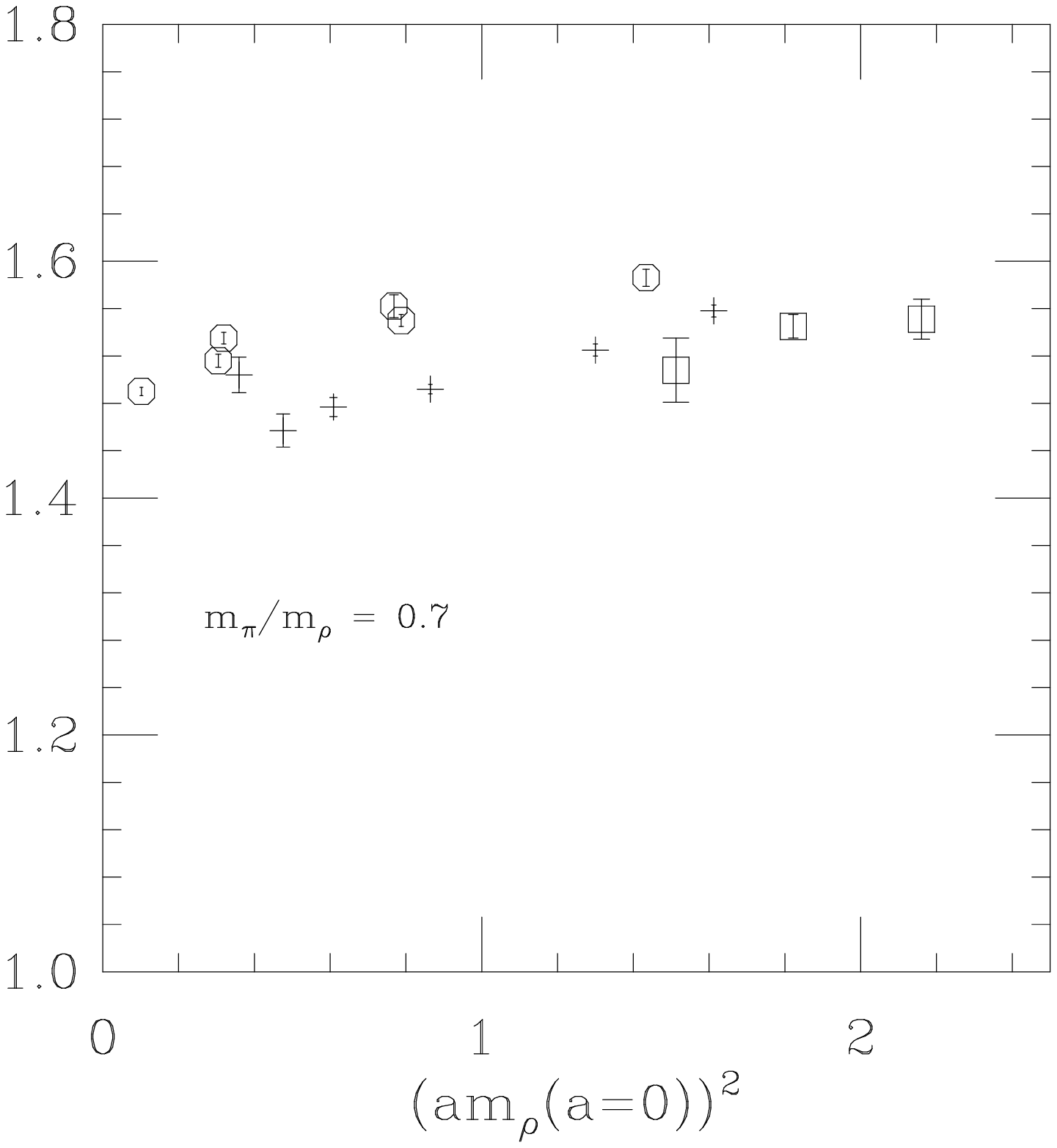}{80mm}
}
\caption{ $m_N/ m_\rho$ vs $(a m_\rho)^2$ at fixed quark mass
(fixed $m_\pi/m_\rho$). }
\label{fig:ratioimp0.7sq}
\end{figure}

Now we turn to tests of quenched QCD for light quarks. 
The two actions which have been most
extensively tested are the S-W action, with and without tadpole improvement,
and an action called the D234(2/3) action, a higher-order variant of
the S-W action  \cite{ALFORD}. Figs. \ref{fig:ratiovsmrhoa96}
and \ref{fig:sommerfigim}, are the analogs of Figs. 
\ref{fig:ratiovsmrhoa} and \ref{fig:sommerfig}.
Diamonds  \cite{UKQCD} and plusses  \cite{SCRI} are S-W actions,
ordinary and tadpole-improved, squares are the D234(2/3) action.
They appear to have about half the scaling violations
as the standard actions but
 they don't remove all scaling violations.
It's a bit hard to quantify the extent of improvement
from these pictures because a chiral extrapolation is hidden in them.
However, one can take one of the ``sections'' of Fig. \ref{fig:fig3combo}
and overlay the new data on it, Fig. \ref{fig:ratioimp0.7}.
It looks like one can double the lattice spacing for an equivalent
amount of scale violation. However, the extrapolation in $a$ is not altogether
clear. Fig. \ref{fig:ratioimp0.7sq} is the same data as Fig.
 \ref{fig:ratioimp0.7}, only plotted vs. $a^2$, not $a$. All of the
actions shown in these figures are supposed to have $O(a^2)$ 
(or better) scaling
violations. Do the data look any straighter in Fig.
\ref{fig:ratioimp0.7sq} than in Fig.  \ref{fig:ratioimp0.7}?

\subsection{The bottom line}
At the cost of enormous effort, one can do fairly high
precision simulations of QCD in the quenched approximation with
standard actions. The actions I have shown you appear to
reduce the amount of computation required for
 pure gauge simulations from supercomputers to
very large work stations, probably  a gain of a few hundreds.
All of the light quark data I showed actually came from supercomputers.
According to Eq. \ref{COST}, a factor of 2 in the lattice spacing
gains a factor of 64 in speed. The cost of either
of the two improved actions I showed is about a factor of 8-10
times the fiducial staggered simulation. Improvement methods
for fermions are a few years less mature than ones for pure gauge
theory, and so the next time you hear a talk about the lattice,
things will have changed for the better (maybe).

\section{SLAC Physics from the Lattice }
One of the major goals of lattice calculations is to provide hadronic
matrix elements which either test QCD or can be used as inputs to
test the standard model. In many cases the lattice numbers have
uncertainties which are small enough that they are interesting to
experimentalists. I want to give a survey of lattice calculations of
matrix elements, and what better way at this summer school, than to recall
 science which was done here at SLAC, as the framework?
\subsection{Generic Matrix Element Calculations}
Most of the matrix elements measured on the lattice are expectation values of
local operators composed of quark and gluon fields.  The mechanical part of
the lattice calculation begins by writing down some Green's
 function which contains the local operator (call it $J(x)$) and somehow
extracting the matrix element. For example, if one wanted
$\langle 0 | J(x) | h\rangle$ one could look at the two-point function
\bee
C_{JO}(t)= \sum_x \langle 0 | J(x,t) O(0,0) | 0 \rangle .
\ee
Inserting a complete set of correctly normalized momentum eigenstates
\bee
1 = {1 \over L^3} \sum_{A, \vec p}{{|A, \vec p \rangle \langle A, \vec p|}
\over {2E_A(p)}} 
\ee
and using translational invariance and going to large $t$ gives
\bee
C_{JO}(t) = e^{-m_A t} {{\langle 0|J|A\rangle \langle A|O|0\rangle}\over
{2m_A}}. 
\ee
A second calculation of
\bee
C_{OO}(t)= \sum_x \langle 0 | O(x,t) O(0,0) | 0 \rangle
= e^{-m_A t} {{|\langle 0|O|A|\rangle|^2}\over
{2m_A}}
\ee
is needed to extract $\langle 0|J|A\rangle$ by fitting  two
correlators with three parameters.
 
Similarly, a matrix element $\langle h | J | h' \rangle$ can be gotten
from
\bee
C_{AB}(t,t') = \sum_x \langle 0 | O_A(t) J(x,t') O_B(0) | 0 \rangle.
\label{PROTME}
\ee
by stretching the source and sink operators $O_A$ and $O_B$ far apart
on the lattice, letting the lattice project out the lightest states,
and then measuring and dividing out $\langle 0 | O_A |h\rangle$
and $\langle 0 | O_B|h\rangle $.
 
These lattice matrix elements are not yet the continuum matrix elements.
The lattice is a UV regulator and changing from the lattice cutoff to a
continuum regulator (like $\overline{MS}$) introduces a shift
\bee
\langle f | O^{cont}(\mu= 1/a) | i \rangle_{\overline{MS}} = a^D
(1+{\alpha_s \over{4\pi}}(C_{\overline{MS}} -C_{latt}) + \dots )
\langle f | O^{latt}(a) | i \rangle  + O(a)+ \dots. 
\label{ZFACTOR}
\ee
The factor $a^D$ converts the dimensionless lattice number to
its continuum result.
The $O(a)$ corrections arise because the lattice
operator might not be the continuum operator:
$df/dx = (f(x+a)-f(x))/a + O(a)$.
The C's are calculable in perturbation theory, and the ``improved
perturbation theory'' described in the last section is often used
to reduce the difference $C_{\overline{MS}} -C_{latt}$.

\subsection{Structure Functions}
In the beginning there was deep inelastic scattering.  The lattice
knows about structure functions through their moments:
\bee
\int_0^1 dx x^{n-2} F_2(x, Q^2) \equiv M_n(Q^2) = \sum c_n^f v_n^{(f)}
\ee
has a representation in terms of matrix elements of fairly complicated
quark (for the nonsinglet structure function) or gluon
bilinears: for quarks
\bee
  \cO ^{(q)}_{\mu_1 \cdots \mu_n} =
       \left(\frac{\mbox{i}}{2}\right)^{n-1} \bar{q}\gamma_{\mu_1}
          \Dd{\mu_2} \cdots \Dd{\mu_n} q
\ee
\bee
\sum_s \langle p,s| \cO^{(f)}_{\{\mu_1\dots\mu_n\}} |p,s\rangle
\simeq v_n^{(f)}(p_{\mu_1} \dots p_{\mu_n} + \dots).
\ee
$D$ is a lattice covariant derivative, which is 
approximated by a finite difference.
The Wilson coefficients $c^{(f)}_{i,n}$
are calculated in perturbation theory and depend on $\mu^2 / Q^2$ as
well as on the coupling constant $g(\mu)$.
The lattice calculation is done by sandwiching the operator
 in Eq. \ref{PROTME}.

\begin{figure}
\centerline{\ewxy{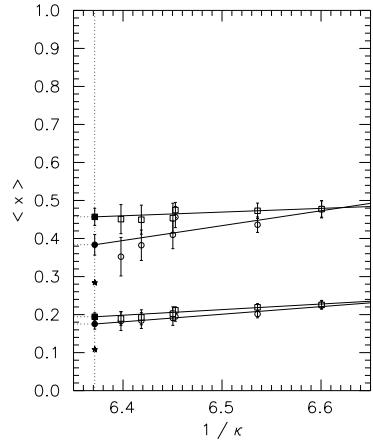}{80mm}
}
\caption{
$\langle x \rangle$ for the proton (MOM scheme) from Ref. 35.
       The circles (boxes) correspond to different choices of lattice operators.
       The upper (lower) band of data represents
       the results for the up (down)-quark distribution.
}
\label{fig:gockx}
\end{figure}

It is presently possible to calculate the two lowest moments of the
proton structure function on the lattice. 
Two groups \cite{GOCKELER,NEGELE} presented results at this
year's lattice conference. Fig. \ref{fig:gockx} shows $\langle x\rangle$
from Ref. \cite{GOCKELER}, a calculation done in quenched approximation. 
In this picture the massless quark limit is the
left edge of the picture. There are several different lattice operators
which serve as discretizations of the continuum operator, and the
figure shows two possibilities.

Unfortunately, the calculation is badly compromised by the quenched approximation. It shows $\langle x\rangle_u=0.38$ and 
$\langle x\rangle_d=0.19$, while in the real world we expect about 0.28
and 0.10 respectively. In the computer there are no sea quarks, and their
momentum is obviously picked up (at least partly) by the valence quarks.

\begin{figure}
\centerline{\ewxy{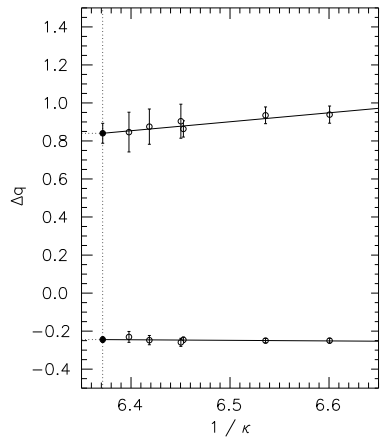}{80mm}
}
\caption{
$\Delta u$ (upper values) and $\Delta d$ (lower values)
         for the proton from Ref. 35.
}
\label{fig:gockx5}
\end{figure}

One nice feature about the lattice calculation is that the spin structure
function can be calculated in essentially the same way; the operator
$\cO$ has an extra gamma-5 in it. Ref. \cite{GOCKELER} computed
$\Delta u =0.84$ and $\Delta d=-0.24$ (in contrast to 0.92 and -0.34
in the real world).
A plot vs quark mass is shown in Fig. \ref{fig:gockx5}.

There is no problem in principle which prevents extending these calculations
to full QCD (with dynamical sea quarks). It will probably be very
 expensive to push beyond the lowest moments.

\subsection{Heavy Quark Physics}
Then there was the November revolution. Twenty years later,
systems with one or more heavy quarks remain interesting objects for study.
The lattice is no exception. Many groups study spectroscopy, decay amplitudes,
form factors... with the  goal of confronting both experiment
and analytic theoretical models.

There are several ways to study heavy quarks on the lattice.
If the quark has infinite mass (the ``static limit'') its
propagator is very simple: the quark is confined to one spatial
location, and as it evolves in time, its color ``twinkles.''
The propagator is just a product of link matrices going forward in time.

One can simulate nonrelativistic quarks directly on the lattice\cite{REF6}.
This has evolved into one of the most successful (and most elaborate)
lattice programs. The idea is to write down lattice actions which are
organized in an expansion of powers of the quark velocity and to
systematically keep all the terms to some desired order. For example,
one might write
\bee
S =  \psi^\dagger [iD_t - {{D^2}\over{2m}} + \vec \mu \cdot \vec B + \dots] \psi
\ee
including kinetic and magnetic moment terms, suitably (and artistically)
discretized. Tadpole-improved perturbation theory is heavily used
to set coefficients.  Figs. \ref{fig:upsil} and \ref{fig:upsiliss}
show the Upsilon spectrum and its hyperfine splittings from various
NRQCD calculations (from a recent summary by Shigemitsu 
\cite{SHIGEMITSU}).

\begin{figure}[t]
\begin{center}
\setlength{\unitlength}{.02in}
\begin{picture}(130,140)(10,930)
\put(15,935){\line(0,1){125}}
\multiput(13,950)(0,50){3}{\line(1,0){4}}
\multiput(14,950)(0,10){10}{\line(1,0){2}}
\put(12,950){\makebox(0,0)[r]{9.5}}
\put(12,1000){\makebox(0,0)[r]{10.0}}
\put(12,1050){\makebox(0,0)[r]{10.5}}
\put(12,1060){\makebox(0,0)[r]{GeV}}
\put(15,935){\line(1,0){115}}
 
 
\multiput(80,1080)(3,0){3}{\line(1,0){2}}
\put(89,1080){\makebox(0,0)[l]{: {\bf Experiment}}}
\put(81,1074){\makebox(0,0)[l]{$\,\Box $ : {\bf Fermilab
$(n_f = 0)$}}}
\put(85,1068){\makebox(0,0)[tl]{\circle{4}}}
\put(89,1068){\makebox(0,0)[l]{: {\bf NRQCD $(n_f = 0)$}}}
\put(85,1062){\makebox(0,0)[tl]{\circle*{4}}}
\put(89,1062){\makebox(0,0)[l]{: {\bf NRQCD $(n_f = 2)$}}}
\put(81,1056){\makebox(0,0)[l]{$\,\Box $ : {\bf SCRI
$(n_f = 2)$}}}
\put(81,1056.7){\makebox(0,0)[l]{$\,+ \;\;$  }}

\put(27,930){\makebox(0,0)[t]{${^1S}_0$}}
\put(25,943.1){\circle{4}}
\put(30,942){\circle*{4}}
\put(33,942){\makebox(0,0){$\Box$}}
\put(22,941.5){\makebox(0,0){$\Box$}}
\put(22,942.5){\makebox(0,0){$+$}}
 
\put(25,1002){\circle{4}}
\put(33,1001){\makebox(0,0){$\Box$}}
\put(33,1001.4){\line(0,1){2.6}}
\put(33,1001.4){\line(0,-1){2.6}}
 
\put(52,930){\makebox(0,0)[t]{${^3S}_1$}}
\put(66,946){\makebox(0,0){1S}}
\multiput(43,946)(3,0){7}{\line(1,0){2}}
\put(50,946){\circle{4}}
\put(55,946){\circle*{4}}
\put(60,945.5){\makebox(0,0){$\Box$}}
\put(45,945.5){\makebox(0,0){$\Box$}}
\put(45,946.5){\makebox(0,0){$+$}}
 
\put(66,1002){\makebox(0,0){2S}}
\multiput(43,1002)(3,0){7}{\line(1,0){2}}
\put(50,1004.1){\circle{4}}
\put(55,1003){\circle*{4}}
\put(55,1004){\line(0,1){1.4}}
\put(55,1002){\line(0,-1){1.4}}
\put(60,1003.5){\makebox(0,0){$\Box$}}
\put(60,1004){\line(0,1){2.7}}
\put(60,1004){\line(0,-1){2.7}}
\put(45,1003.5){\makebox(0,0){$\Box$}}
\put(45,1004.3){\makebox(0,0){$+$}}
\put(45,1003.9){\line(0,1){4}}
\put(45,1003.9){\line(0,-1){4}}

\put(66,1036){\makebox(0,0){3S}}
\multiput(43,1036)(3,0){7}{\line(1,0){2}}
\put(50,1060){\circle{4}}
\put(50,1060){\line(0,1){11}}
\put(50,1060){\line(0,-1){11}}
\put(55,1039.1){\circle*{4}}
\put(55,1039.1){\line(0,1){7.2}}
\put(55,1039.1){\line(0,-1){7.2}}
 
\put(92,930){\makebox(0,0)[t]{${^1P}_1$}}
 
\put(106,990){\makebox(0,0){1P}}
\multiput(83,990)(3,0){7}{\line(1,0){2}}
\put(90,987.6){\circle{4}}
\put(95,989){\circle*{4}}
\put(100,989.5){\makebox(0,0){$\Box$}}
\put(87,989.5){\makebox(0,0){$\Box$}}
\put(87,990.3){\makebox(0,0){$+$}}

\put(106,1026){\makebox(0,0){2P}}
\multiput(83,1026)(3,0){7}{\line(1,0){2}}
\put(90,1038.7){\circle{4}}
\put(95,1023){\circle*{4}}
\put(95,1023){\line(0,1){7.2}}
\put(95,1023){\line(0,-1){7.2}}

\put(130,1025){\makebox(0,0){1D}}
\put(120,930){\makebox(0,0)[t]{${^1D}_2$}}
\put(120,1019.2){\circle{4}}
\put(120,1019.2){\line(0,1){6}}
\put(120,1019.2){\line(0,-1){6}}
 
\end{picture}
\end{center}

\caption{{  $\Upsilon$ Spectrum }
\label{fig:upsil}
}
\end{figure}
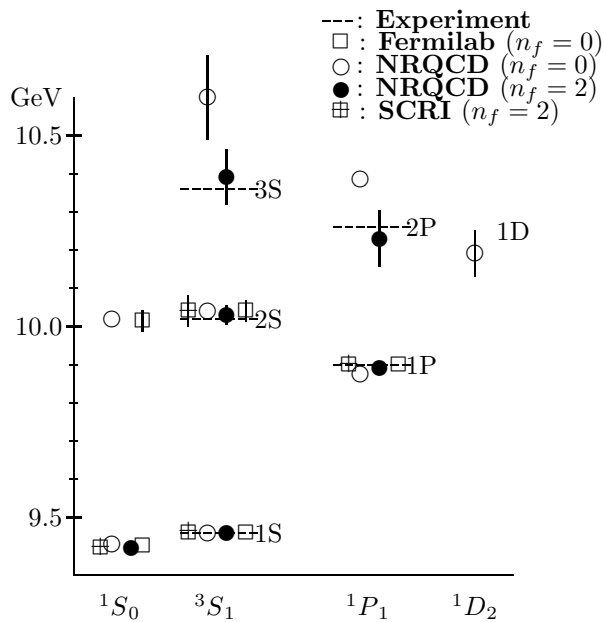

\begin{figure}
\begin{center}
\setlength{\unitlength}{.02in}
\begin{picture}(100,80)(15,-50)
 
\put(15,-50){\line(0,1){80}}
\multiput(13,-40)(0,20){4}{\line(1,0){4}}
\multiput(14,-40)(0,10){7}{\line(1,0){2}}
\put(12,-40){\makebox(0,0)[r]{$-40$}}
\put(12,-20){\makebox(0,0)[r]{$-20$}}
\put(12,0){\makebox(0,0)[r]{$0$}}
\put(12,20){\makebox(0,0)[r]{$20$}}
\put(12,30){\makebox(0,0)[r]{MeV}}
\put(15,-50){\line(1,0){100}}

 
\multiput(28,0)(3,0){7}{\line(1,0){2}}
\put(50,2){\makebox(0,0)[t]{$\Upsilon$}}
\put(35,0){\circle{4}}
\put(40,0){\circle*{4}}
\put(45,-0.5){\makebox(0,0){$\Box$}}
\put(30,-0.5){\makebox(0,0){$\Box$}}
\put(30,0.5){\makebox(0,0){$+$}}
 
\put(48,-34){\makebox(0,0)[t]{$\eta_b$}}
\put(35,-29.9){\circle{4}}
\put(40,-39){\circle*{4}}
\put(40,-39){\line(0,1){2}}
\put(40,-39){\line(0,-1){2}}
\put(43,-35.3){\makebox(0,0){$\Box$}}
\put(32,-43){\makebox(0,0){$\Box$}}
\put(32,-42.3){\makebox(0,0){$+$}}
\put(32,-43){\line(0,1){6}}
\put(32,-43){\line(0,-1){6}}
 
\put(63,-5){\makebox(0,0)[l]{$h_b$}}
\put(70,-1.8){\circle{4}}
\put(75,-2.9){\circle*{4}}
\put(75,-2.9){\line(0,1){1.2}}
\put(75,-2.9){\line(0,-1){1.2}}
\put(78, 0.){\makebox(0,0){$\Box$}}
\put(68, 0.){\makebox(0,0){$\Box$}}
\put(68, 0.9){\makebox(0,0){$+$}}
 
\multiput(90,-40)(3,0){7}{\line(1,0){2}}
\put(110,-40){\makebox(0,0)[l]{$\chi_{b0}$}}
\put(97,-25.1){\circle{4}}
\put(102,-34){\circle*{4}}
\put(102,-34){\line(0,1){5}}
\put(102,-34){\line(0,-1){5}}
\put(107,-33){\makebox(0,0){$\Box$}}
\put(107,-33){\line(0,1){18}}
\put(107,-33){\line(0,-1){17}}
 
\multiput(90,-8)(3,0){7}{\line(1,0){2}}
\put(110,-8){\makebox(0,0)[l]{$\chi_{b1}$}}
\put(97,-8.6){\circle{4}}
\put(102,-7.9){\circle*{4}}
\put(102,-7.9){\line(0,1){2.4}}
\put(102,-7.9){\line(0,-1){2.4}}
\put(105,-12){\makebox(0,0){$\Box$}}
\put(105,-12){\line(0,1){14}}
\put(105,-12){\line(0,-1){16}}
 
\multiput(90,13)(3,0){7}{\line(1,0){2}}
\put(110,13){\makebox(0,0)[l]{$\chi_{b2}$}}
\put(97,10.2){\circle{4}}
\put(102,11.5){\circle*{4}}
\put(102,11.5){\line(0,1){2.4}}
\put(102,11.5){\line(0,-1){2.4}}

\end{picture}
\end{center}
 \caption{ $\Upsilon$ Spin Splittings: Symbols have the
same meaning as in Figure 25.}
\label{fig:upsiliss}
\end{figure}
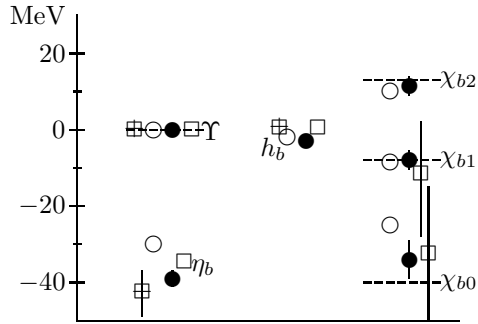

The main shortcoming of nonrelativistic QCD is of course that when the
quark mass gets small the nonrelativistic approximation breaks down.
For charmonium $v/c \simeq 0.3$, so the method is less safe for this
system than for the Upsilon.

Finally, one can take relativistic lattice quarks and just make the mass
heavy. If the quark mass gets too heavy ($ma \simeq 1$)
lattice artifacts dominate the calculation. For Wilson fermions,
the dispersion relation breaks down: $E(p) \simeq m_1 + p^2/2m_2$
where $m_2 \neq m_1$. The magnetic moment is governed by its
own different mass, too.

Another signal of difficulty is that all these  formulations have
their own pattern of scale violations. That is, nonrelativistic quarks
and Wilson quarks approach their $a\rightarrow 0$ limits differently.
This is often described in the literature by the statement that ``the lattice
spacing is different for different observables.'' For example, in one
data set\cite{ALPH}, the inverse lattice spacing (in MeV) is given as
2055 MeV from fitting the heavy $q \bar q$ potential, 2140 MeV from the
rho mass, 1800 MeV from the proton mass, and 2400 MeV from the Upsilon
 spectrum. These simulations are just sitting in the middle of figures
like Fig. \ref{fig:ratiovsmrhoa} with only one point, trying to
guess where the left hand edge of the picture will be.
This is a problem for calculations of B meson and baryon spectroscopy,
where the heavy quarks might be treated nonrelativistically and
the light quarks are relativistic. What observable should be used
to set the overall scale?\cite{SCRIB}

One of the major uses of heavy quark systems by the lattice community
is to try to calculate the strong coupling constant at $Q^2=M_Z^2$. This
topic deserves its own section.

\subsection{$\alpha_s(M_Z)$}
Now we are at the SLC and LEP.
For some time now there have been claims that physics at the Z pole
hints at a possible breakdown in the standard model  \cite{SHIFMAN}.
A key question in the discussion is whether or not
 the value of $\alpha_{\overline{MS}}$ inferred from the decay width of the
Z is anomalously high relative to other determinations of the strong
coupling (which are usually measured at lower $Q$ and
 run to the Z pole).

The most recent analysis of $\alpha_s(M_Z)$ I am aware of is due to
Erler and Langacker  \cite{PDGALPHA}.
Currently, $\alpha_{\overline{MS}}^{lineshape}=  0.123(4)(2)(1)$ for the standard
model Higgs mass range, where the first/ second/ third uncertainty
is from inputs/ Higgs mass/ 
estimate of $\alpha_s^4$ terms. The central Higgs mass
is assumed to be 300 GeV, and the second error is for $M_H=1000$  GeV (+),
 60 GeV (-).
For the SUSY Higgs mass range (60-150 GeV), one has the lower value 
$\alpha_{\overline{MS}}=.121(4)(+1-0)(1)$. A global fit to all data gives
0.121(4)(1). Hinchcliffe in the same compilation quotes a global average
of 0.118(3).

The lattice can contribute to this question by predicting 
$\alpha_{\overline{MS}}$ from low energy physics. The basic idea is simple:
The lattice is a (peculiar) UV cutoff. A lattice mass $\mu = Ma$ plus
an experimental mass $M$ give a lattice spacing $a = \mu/M$ in fm.
If one can measure some quantity related to $\alpha_s$ at a scale
$Q\simeq 1/a$, one can then run the coupling constant out to the Z.

\begin{figure}
\centerline{\ewxy{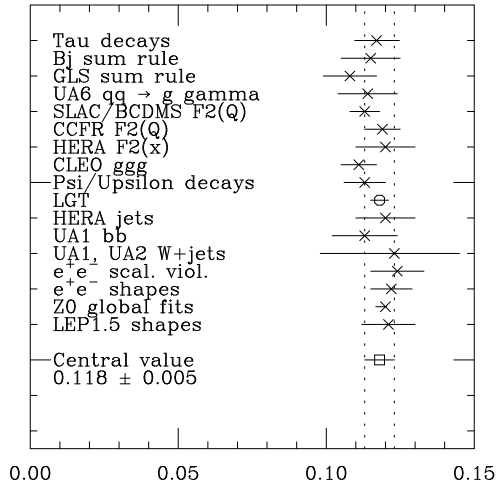}{80mm}
}
\caption{ Survey of $\alpha_{\overline{MS}}(M_Z)$ from Ref. 42.}
\label{fig:allalpha}
\end{figure}

The best (recent) lattice number, from Shigemitsu's Lattice 96
summary talk  \cite{SHIGEMITSU}, is
\bee
\alpha_{\overline{MS}}(M_Z) = 0.1159(19)(13)(19)
\ee
where the first error includes both statistics and estimates of
discretization errors, the second is due to uncertainties from the
dynamical quark mass, and the third is from conversions of conventions.
The lattice number is about one standard deviation below the pure
Z-physics number. Lattice results are compared
to other recent determinations of $\alpha_{\overline{MS}}(Z)$
 in Fig. \ref{fig:allalpha}, a figure provided by P. Burrows \cite{BURROWS}.

Two ways of calculating $\alpha_s(M_Z)$ from lattice have been proposed:
The first is the ``small loop method'' 
  \cite{AIDA}. This method uses the ``improved
perturbation theory'' described in  Chapter 2:
One assumes that a version of perturbation theory can describe the behavior
of short distance objects on the lattice: in particular, that the
plaquette can be used to define $\alpha_V(q=3.41/a)$. With typical
lattice spacings now in use, this gives the coupling at a momentum 
$Q_0=8-10$ GeV.  One then converts the coupling to $\alpha_{\overline{MS}}$
and runs out to the Z using the (published) three-loop beta function
 \cite{ROD}.

Usually, the lattice spacing is determined from the mass splittings
of heavy $Q \bar Q$ states. This is done because the mass differences between
physical heavy quark states are nearly independent of the quark mass--
for example, the S-P mass splitting of the $\psi$ family is about 460
MeV, and it is about 440 MeV for the $\Upsilon$. A second reason is that
the mass splitting is believed to be much less sensitive to sea
quark effects than light quark observables, and one can estimate
the effects of sea quarks through simple potential models.
The uncertainty in the lattice spacing is three to five per cent,
but systematic effects are much greater (as we will see below).

The coupling constant comes from Eq. \ref{PLAQALPHA}. The plaquette can be
measured to exquisite accuracy (0.01 per cent is not atypical) and so the 
coupling constant is known essentially without error. However, the
scale of the coupling is uncertain (due to the lattice spacing).

The next problem is getting from lattice simulations, which are done with
$n_f=0$ (quenched) or $n_f=2$ (but unphysical sea quark masses)
to the real world of $n_f=3$. Before simulations with dynamical fermions
were available, the translation was done by running down in $Q$ to
a ``typical gluonic scale'' for the psi or the upsilon (a few hundred MeV)
and then matching the coupling to the three-flavor coupling (in the spirit
of effective field theories). 
This produced a rather low $\alpha_s\simeq 0.105$.
Now we have simulations at $n_f=2$ and can do better.
Recall that in lowest order
\bee
{1 \over \alpha_s} = \big( {{11 - {2 \over 3}n_f} \over {4\pi}}  \big) 
\ln {Q^2 \over \Lambda^2}
\label{ALINV}
\ee
One measures $1/\alpha_s$ in two simulations, one quenched, the other
at $n_f=2$, runs one measurement in $Q$ to the $Q$ of the other,
then extrapolates $1/\alpha$ linearly in $n_f$ to $n_f=3$.
Then one can convert to $\overline{MS}$ and run away.

\begin{figure}
\centerline{\ewxy{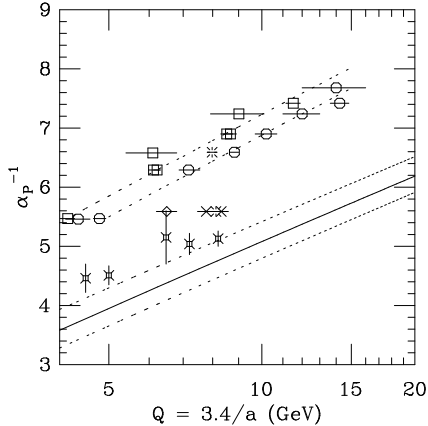}{80mm}
}
\caption{ Survey of $\alpha_{\overline{MS}}(Q)$ at the scale where lattice
simulations are actually done.}
\label{fig:alphap2js}
\end{figure}

Pictures like Fig. \ref{fig:allalpha} are not very useful
when one wants to get a feel for the errors inherent in the lattice
calculation. Instead, let's run our expectations for $\alpha_s(M_Z)$
down to the scale where the lattice simulations are done, and compare.
Fig. \ref{fig:alphap2js} does that. The squares are the results of
simulations of charmed quarks and the octagons are from bottom
quarks, both with $n_f=0$. The crosses and diamond are $n_f=2$
bottom and charm results. 
(The bursts show upsilon data when the 1S-2S mass difference gives a 
lattice spacing.) 
Note the horizontal error bars on the lattice data.
Finally, the predicted $n_f=3$ coupling $\alpha_P$ is shown
in the fancy squares, with error bars now rotated because the
convention is to quote an error in $\alpha_s$.  The lower three lines
in the picture (from top to bottom) 
are $\alpha_{\overline{MS}}(M_Z)=0.118$, 0.123, and 0.128
 run down and converted to the lattice prescription.

The two top lines are predictions for how quenched $\alpha$ should run.

Now for the bad news. All of the $n_f=2$ data shown here were actually 
run on the same set of configurations. The bare
couplings are the same, but the
lattice spacings came out different. 
What is happening is that we are taking  calculations at some lattice
spacing and inferring a continuum numbers from them, but the lattice
predictions have scale violations which are different.
(The $\Upsilon$ calculations use nonrelativistic quarks, the
$\psi$ calculations use heavy Wilson quarks.)
Notice also that the bottom and
charm quenched lattice spacings are different. This discrepancy
is thought to be a failure of the quenched approximation: the characteristic
momentum scale for binding in the $\psi$ and $\Upsilon$ are different,
and because $n_f$ is not the real world value $\alpha$ runs incorrectly
between the two scales.
Said differently, in the quenched approximation, the spectrum of
heavy quark bound states is different from the real world.

There is a second method of determining a running coupling constant which
actually allows one to see the running over a large range of scales.
It goes by the name of the ``Schr\"odinger functional,''
 (referring to the
fact that the authors study QCD in a little box with specified
boundary conditions) but
``coupling determined by varying the box size'' would be a more descriptive
title. It has been applied to quenched QCD but has not yet been
extended to full QCD  \cite{SCHQCD}, and so it is not
yet had an impact on phenomenology.
This calculation
does not use perturbation theory overtly.  
For a critical comparison of the two methods,
see \cite{WEISZAL} .

\subsection{Glueballs}
I have been hearing people talk about glueballs from psi decay
for almost twenty years.
Toki \cite{TOKI} has summarized the experimental situation for
glueballs.
What do theorists expect for a spectrum? 
The problem is that any non-lattice model requires making
uncontrolled approximations to get any kind of an answer:
there are no obvious zeroth order solutions with small
expansion  parameters.
The lattice is the 
only game in town for a first-principles calculation.

People have been trying to measure the masses of the lightest
glueballs (the scalar and the tensor)
using lattice simulations for many years. 
The problem has proven to be very hard, for several reasons.

Recall how we measure a mass from a correlation function (Eq. \ref{CORRFN}).
The problem with the scalar glueball is that
 the operator $O$ has nonzero vacuum expectation
value, and the correlation function approaches a constant at large $t$:
\bee
\lim_{t \rightarrow \infty}C(t) \rightarrow |\langle 0|O|\vec p = 0\rangle|^2
\exp(-mt) + |\langle 0|O| 0\rangle|^2 .
\ee
The statistical fluctuations on $C(t)$ are given by Eq. \ref{STDEV}
 and we find after a short calculation that
\bee
\sigma \rightarrow {C(0) \over \sqrt{N}} . \ee
Thus the signal to noise ratio collapses at large $t$ like $\sqrt{N} \exp(-mt)$
due to the constant term.
 
A partial cure for this problem is a good trial wave function $O$.
While in principle the plaquette itself could be used, it is so
dominated by ultraviolet fluctuations that it does not produce a good signal.
Instead, people invent ``fat links'' which average the gauge field
over several lattice spacings, and then make interpolating fields
which are closed loops of these fat links. The lattice glueball is
a smoke ring.

\begin{figure}
\centerline{\ewxy{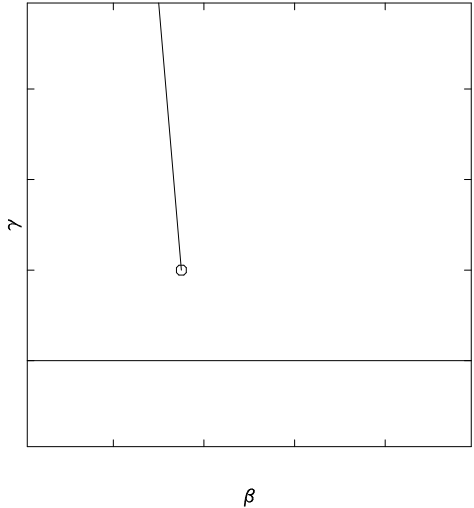}{80mm}
}
\caption{Phase transitions in the fundamental-adjoint plane.}
\label{fig:fundadj}
\end{figure}

The second problem is that lattice actions can have phase transitions
at strong or intermediate coupling, which have nothing to do
with the continuum limit, but mask continuum behavior \cite{BHAN}.
 As an example
of this, consider the gauge group $SU(2)$, where a  link
matrix can be parameterized as
 $U = 1\cos \theta + i \vec \sigma \cdot \vec n \sin \theta$,
so ${\rm Tr}U = 2 \cos \theta$. Now consider a generalization
of the Wilson action $-S = \beta {\rm Tr}U + \gamma({\rm Tr}U)^2$.
(this is a mixed fundamental-adjoint representation action).
At $\gamma \rightarrow \infty$ ${\rm Tr}U \rightarrow \pm 1$ and the
gauge symmetry is broken down to $Z(2)$. But $Z(2)$ gauge theories have
a first order phase transition. First order transitions are stable under
perturbations, and so the phase diagram of this theory, shown
in Fig. \ref{fig:fundadj}, has a line of first order transitions 
which terminate in a second order point. At the second order point some
state with scalar quantum numbers becomes massless. However, now
imagine that you are doing Monte Carlo along the $\gamma=0$ line, that is,
with the Wilson action. When you come near the critical point, any operator
which couples to a scalar particle (like the one you are using to see the 
scalar glueball) will see the nearby transition and the lightest mass
in the scalar channel will shrink. Once you are past the point of
closest approach, the mass will rise again. Any scaling test which ignores
the nearby singularity will lie to you.

\begin{figure}
\centerline{\ewxy{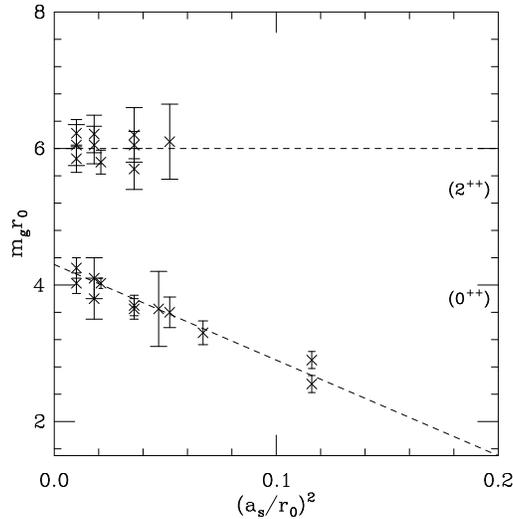}{80mm}
}
\caption{Glueball mass vs $r_0$ with the Wilson action, from a summary
picture in Ref. 49.}
\label{fig:gbr0small}
\end{figure}

This scenario has been mapped out for $SU(3)$, and the place of closest
approach is at a Wilson coupling corresponding to a lattice spacing
of 0.2 fm or so, meaning that very small lattice spacings are needed
before one can extrapolate to zero lattice spacing.
A summary of the situation is shown in Fig. \ref{fig:gbr0small}
 \cite{PEARDON}.
Here the quantity $r_0$ is the ``Sommer radius''  \cite{SOMMER},
defined through the heavy quark force, $F(r)= - dV(r)/dr$,
 by $r_0^2F(r_0)= -1.65$.
In the physical world of three colors and four flavors, $r_0 = 0.5$ fm.

Finally, other arguments suggest that a small lattice spacing or
a good approximation to an  action on an RT are needed to for
glueballs:  the physical diameter of the glueball, as inferred from
the size of the best interpolating field, is small, about 0.5 fm.
 Sh\"afer and Shuryak  \cite{SS} have argued that the small size is due to
instanton effects.
Most lattice actions do bad things to instantons 
at large lattice spacing\cite{INSTANTON12}.

Two big simulations have carried calculations of the glueball mass
close to the continuum limit: the UKQCD collaboration  \cite{UKQCDGB}
and a collaboration at IBM which built its own computer  \cite{GF11}.
(The latter group is the one with the press release last December
announcing the discovery of the glueball.)
Their predictions in MeV are different and they each favor a different
experimental candidate for the scalar glueball (the one which is
closer to their prediction, of course). It is a useful object lesson
because both groups say that their lattice numbers agree before
extrapolation, but they extrapolate differently to $a=0$.

The UKQCD group sees that the ratio $m(0^{++})/\surd\sigma$ can be
well fitted with  a form   $b + c a^2 \sigma$
($\sigma$ is the string tension) and a fit of this form to the lattice
data of both groups gives
$ m(0^{++})/\surd\sigma = 3.64 \pm 0.15$.
To turn this into MeV we need $\sigma$ in MeV units. One way
is to take $m_{\rho}/\surd\sigma$ and extrapolate that
to $a=0$ using $ b + c a \surd\sigma$. 
Averaging and putting 770 MeV for $m_{\rho}$ one gets
$\surd\sigma = 432 \pm 15$ MeV, which is  consistent
with the usual estimate (from extracting the string tension
from the heavy quark potential)
of about 440 MeV.
Using the total average they get 
$m(0^{++}) = 1572 \pm 65 \pm 55$ MeV
where the first error is statistical and the second comes
from the scale. 

The IBM group, on the other hand, notices that $m_\rho a$ and $m_\phi a$
scale asymptotically, use the phi mass to predict 
quenched $\Lambda_{\overline{MS}}$,
then extrapolate $m(0^{++})/\Lambda = A + B(a \Lambda)^2$. They get
1740(41) MeV from their data,  when they analyze UKQCD data, they
get 1625(94) MeV, and when they combine the data sets, they get 1707(64)
MeV.

A neutral reporter could get hurt here. It seems to me that the lattice  prediction for the scalar glueball
is $1600 \pm 100 $ MeV, and that there are two experimental candidates
for it, the $f_0(1500)$ and the $f_J(1710)$. 

Masses are not the end of the story. The IBM group has done two 
interesting recent calculations related to glueballs, which strengthen their
claim that the $f_J(1710)$ is the glueball.

\begin{figure}
\centerline{\ewxy{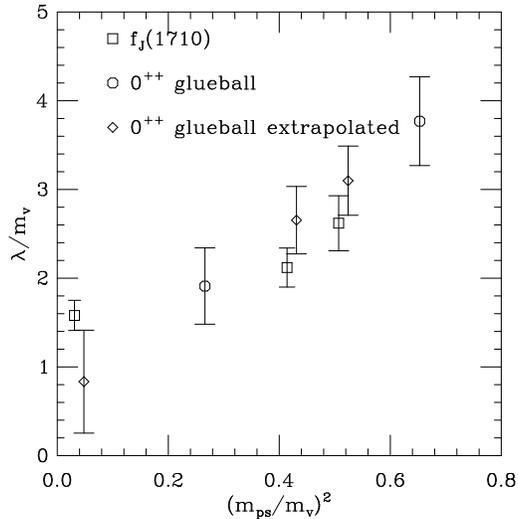}{80mm}
}
\caption{Scalar glueball decay couplings from Ref. 54.}
\label{fig:gbdecay}
\end{figure}

The first one  of them  \cite{GF11DE} was actually responsible for the
press release. It is a calculation of the decay width of the
glueball into pairs of pseudoscalars. This is done by computing
an unamputated three point function on the lattice, with an
assumed form for the vertex, whose magnitude is fitted. The result
 is shown in Fig. \ref{fig:gbdecay}. 
The octagons are the results of the simulation and
the diamonds show interpolations in the quark mass.
The ``experimental'' points (squares) are from
a partial wave analysis of isoscalar scalar resonances by Longacre
and Lindenbaum  \cite{LL}.

The response of a member of the other side is that
the slope of the straight line that one would put through
the three experimental points is barely, if at all, compatible with
the slope of the theoretical points. Since they argue
theoretically for a straight line, the comparison
of such slopes is a valid one. 

If one of the experimental states is not a glueball, it is likely
to be a ${}^3P_0$ orbital excitation of quarks.
 Weingarten and Lee  \cite{GF11P0} are computing the mass
of this state on the lattice and argue that it is lighter than
1700 MeV; in their picture the $f_0(1500)$ is an $s \bar s$ state. 
I have now said more than I know and will just refer you to
  recent discussions of the question  \cite{GBFIGHT}.

Both groups predict that the $2^{++}$ glueball is at about 2300 MeV.

\begin{figure}
\centerline{\ewxy{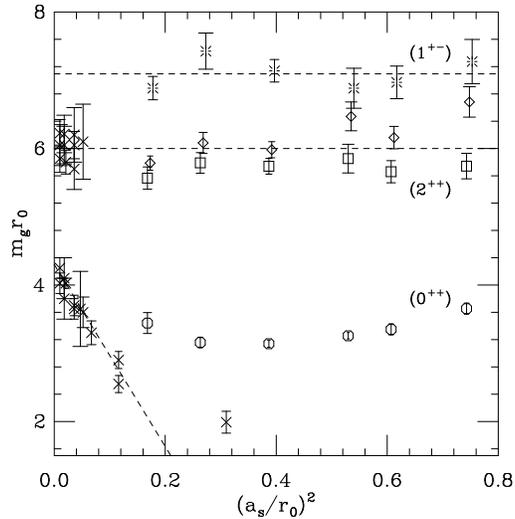}{80mm}
}
\caption{Glueball mass vs $r_0$ from Ref. 49, including large
lattice spacing data.}
\label{fig:gbr0}
\end{figure}

Can ``improved actions'' help the situation? Recently, Peardon and Morningstar
 \cite{PEARDON}
implemented a clever method for beating the exponential signal-to-noise ratio:
make the lattice spacing smaller in the time direction than in the space direction. Then the signal, which falls like $\exp (-m a_t L_t)$ after
$L_t$ lattice spacings, dies more slowly because $a_t$ is reduced.
Their picture of the glueball mass vs $r_0$. is shown in Fig. \ref{fig:gbr0}.
They are using the tadpole-improved L\"uscher-Weisz action.
The pessimist notes the prominent dip in the middle of the curve; this
action also has a lattice-artifact transition (somewhere); the optimist
notes that the dip is much smaller than for the Wilson action and
then the pessimist notes that there is no Wilson action data at large
lattice spacing to compare. I think the jury is still out.

\subsection{The B Parameter of the B Meson}
And finally we are at BaBar. $\bar B-B$ mixing is parameterized by the
ratio
\bee
x_d = {{(\Delta M)_{b \bar d}}\over {\Gamma_{b \bar d}}}
= 
\tau_{b \bar d}{{G_F^2}\over{6\pi^2}}\eta_{QCD}F\big({{m_t^2}\over{m_W^2}}\big)
 |V_{tb}^*V_{td}|^2 
 b(\mu) \{ {3\over 8}\langle \bar B| \bar b \gamma_\rho(1-\gamma_5) d
\bar b \gamma_\rho(1-\gamma_5) d | B \rangle  \}
\label{BIGEQ}
\ee
Experiment is on the left; theory on the right. Moving into the
long equation from the left, we see many known (more or less) parameters
from phase space integrals or perturbative QCD calculations, then
a combination of CKM matrix elements, followed by a four quark hadronic
matrix element \cite{ROSNER}. We would like to extract the CKM
matrix element from the measurement of $x_d$ (and its strange partner
$x_s$). To do so we need to know the value of the
 object in the curly brackets, defined as $3/8 M_{bd}$
and  parameterized as $m_B^2 f_{B_d}^2 B_{b_d}$ where $B_{b_d}$ is the
so-called B-parameter, and $f_B$ is the B-meson decay constant
\bee
\langle 0 | \bar b \gamma_0 \gamma_5 d | B \rangle =
f_Bm_B.
\ee
Naive theory, which is expected to work well for the B system, suggests that
$B_B=1$ to good accuracy.
Of course, the stakes are high and a good determination of $M_{bd}$
is needed to test the standard model. The lattice can do just that.

In Eq. \ref{BIGEQ}  $b(\mu)$, the coefficient which runs the
effective interacion down from the W-boson scale to the QCD scale $\mu$,
and the matrix element $M(\mu)$ both depend on the QCD scale,
and one often sees the renormalization group invariant quantities
$\hat M_{bd} = b(\mu) M_{bd}(\mu)$ or $\hat B_{bd} = b(\mu) B_{bd}(\mu)$
quoted in the literature.

Decay constants  probe very simple properties of the wave function: in the
nonrelativistic quark model
\bee
f_M = {{\psi(0)}\over {\sqrt{m_M}}} 
\ee
where $\psi(0)$ is the $\bar q q$ wave function at the origin.
For a heavy quark ($Q$) light quark ($q$) system $\psi(0)$ should become
independent of the heavy quark's mass as the $Q$ mass goes to infinity, and
in that limit one can show in QCD that $\sqrt{m_M}f_M$ approaches a constant.

One way to compute the decay constant is to put a light quark and a
 heavy quark on the lattice and let them propagate.
It is difficult to calculate $f_B$ directly on present day lattices
with relativistic lattice fermions
because the lattice spacing is much greater than the $b$ quark's
Compton wavelength (or the UV cutoff is below $m_b$).  In this limit
the $b$ quark is strongly affected by lattice artifacts as it propagates.
However, one can
make $m_b$ infinite on the lattice and determine the combination
$\sqrt{m_B} f_B$ in the limit.
 Then one can extrapolate
down to the $B$ mass and see if
 the two extrapolations up and down give the
same result.
(Nonrelativistic $b$ quarks can solve this problem in principle, but
the problem of setting the lattice spacing between light and nonrelativistic
quarks has prevented workers from quoting a useful decay constant from these
simulations.)

Among the many lattice decay constant calculations, the one of
Ref. \cite{BERNARD} stands out in my mind for being  the most complete.
These authors did careful quenched simulations at many values of the lattice
spacing, which allows one to extrapolate to the continuum limit by brute
force. They have also done a less complete set of simulations which
include light dynamical quarks, which should give some idea of the
accuracy of the quenched approximation.

The analysis of all this data is quite involved. One begins with
a set of lattice decay constants measured in lattice units, from simulations
done with heavy quarks which are probably too light and light quarks which
are certainly too heavy. One has to interpolate or extrapolate 
the heavy quark masses to their real world values, extrapolate 
the light quarks down in mass to their physical values, and finally
try to extrapolate to $a\rightarrow 0$. It is not always obvious how to
do this.
Complicating everything
are the lattice artifacts in the fermion and gauge actions, and the
lattice-to-continuum renormalization factors as in Eq. \ref{ZFACTOR}.

\begin{figure}
\centerline{\ewxy{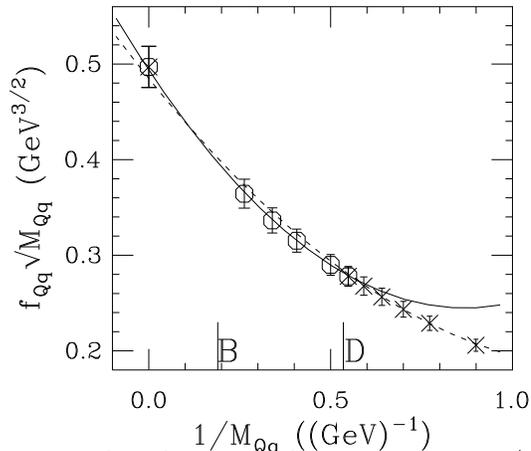}{80mm}
}
\caption{
Pseudoscalar meson decay constant vs $1/M$, from Ref. 59.}
\label{fig:frootm}
\end{figure}

\begin{figure}
\centerline{\ewxy{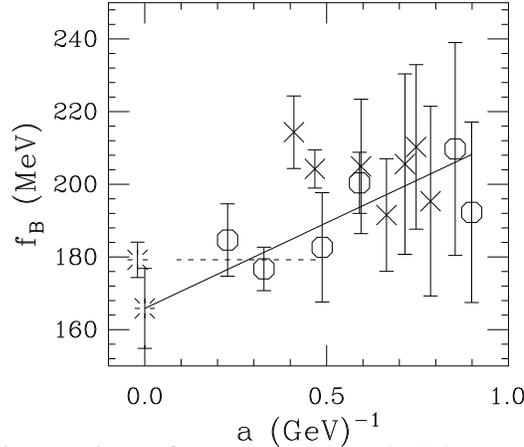}{80mm}
}
\caption{ $f_B $ vs. $a$ from Ref. 59.  Octagons are quenched data;
crosses, $N_F=2$.  The solid line is a linear fit to all
quenched points;  the dashed line is a a constant
fit to the three quenched points with $a<0.5$ GeV${}^{-1}$.
The extrapolated values at $a=0$ are indicated by
bursts.
The scale
is set by $f_\pi=132$ MeV throughout.
}
\label{fig:fb}
\end{figure}

The (still preliminary) results of Ref. \cite{BERNARD} are shown in Figs.
\ref{fig:frootm} and \ref{fig:fb}. The $N_f=2$ dynamical
fermion data in Fig. \ref{fig:fb} have moved
around a bit in the past year and may not have settled down yet.

The numerical results of Ref. \cite{BERNARD} are:
\begin{eqnarray*}
& f_B  =  166(11)(28)(14)  \ & f_D  =   196(9)(14)(8) \\
& f_{B_s}  =  181(10)(36)(18) \   & f_{D_s}  =   211(7)(25)(11) \\
& {f_{B_s}\over f_B}  =   1.10(2)(5)(8) \ & {f_{D_s}\over f_D}  =
1.09(2)(5)(5)
\end{eqnarray*}
where the first error includes statistical
errors and systematic effects of changing fitting ranges; the second, other
errors
within the quenched approximation; the third, an estimate
of the quenching error. Decay constants are in MeV.

Note that the error bars for the B system are small enough to be
phenomenologically interesting.  The Particle Data Group's \cite{PDG}
 determination
of CKM matrix elements, which does not include this data,
 says, ``Using $\hat B_{B_d} f_{B_d}^2 =
(1.2 \pm 0.2)(173 \pm 40  \ {\rm MeV})^2$...,
$ |V_{tb}^*V_{td}|^2=0.009 \pm 0.003$, where the error bar comes primarily from
the theoretical uncertainty in the hadronic matrix elements.''

Can we trust these numbers? Lattice calculations have been predicting
$f_{D_s} \simeq 200$ MeV for about eight years. 
The central values have changed very little, while the uncertainties
have decreased.
So far four experiments have reported measurements of this quantity.
The most recent is Fermilab E653 Collaboration \cite{E653}
       $f_{D_s} = 194(35)(20)(14)$ MeV.
The older numbers, with bigger errors, were 
       238(47)(21)(43)  from  WA75 (1993) \cite{WA75};
       344(37)(52)(42)  from  CLEO  (1994) \cite{CLEO};
       430 (+150 -130)(40) from  BES (1995) \cite{BES}.

Now back to the mixing problem. On the lattice, one could measure
the decay constants and $B$ parameter separately and combine them after
extrapolation, or measure $M$ directly and extrapolate it.
In principle the numbers should be the same, but in practice they will not
be.

\begin{figure}
\centerline{\ewxy{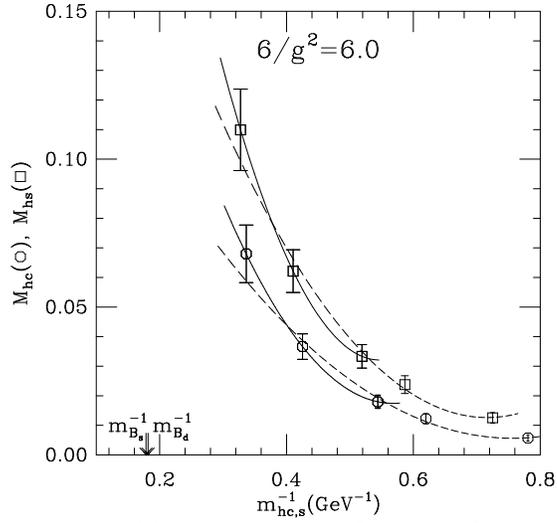}{80mm}
}
\caption{
$\MC$ (octagons) and $\MHS$ (squares) as a
                function of the inverse
                heavy-down(strange) meson mass, at $\beta=6.0$. The
                dashed line shows the effect of the lightest points on the fit.}
\label{fig:fig3blum}
\end{figure}

A recent calculation illustrates this point \cite{BLUM}.
The authors computed the four fermion
operators $M_{bs}$ and $M_{bd}$ directly on the lattice.
Fig. \ref{fig:fig3blum} shows the behavior of $M$ as a function of 
hadron mass at one of their lattice spacings.

\begin{figure}
\centerline{\ewxy{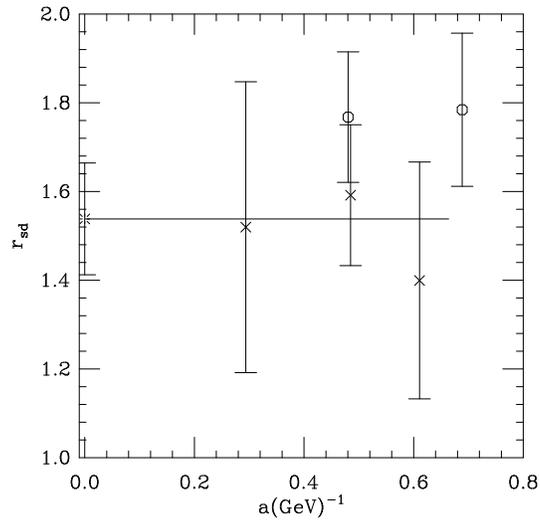}{80mm}
}
\caption{
        The SU(3) flavor breaking ratio $\MS/\MB$
                versus the lattice spacing $a$. The points denoted by crosses
                were used in the fit (solid line). 
 The burst shows the extrapolation to $a=0$.
}
\label{fig:fig4blum}
\end{figure}

The ratio $r_{sd} = M_{bs}/M_{bd}$ is presumably much less sensitive
to lattice spacing or to quark mass extrapolation. The authors' result
for the lattice spacing dependence of this ratio
 is shown in Fig. \ref{fig:fig4blum}, along with an extrapolation to
zero lattice spacing. They find $1.54\pm .13\pm .32$ from their direct
method, compared to $r_{sd}\approx 1.32\pm .23$ from separate extrapolations
of the decay constants and the $B$ parameter. 
(They measure equal B parameters for strange and nonstrange B mesons,
$B(\mu)=1.02(13)$ for $\mu=2$ GeV.)

It looks like
$SU(3)$ breaking is fairly large, and if that is so, it looks like
the parameter $x_s$, the strange analog of Eq. \ref{BIGEQ}, might
be about 20, unmeasurably large \cite{FLYNN}.

\section{Conclusions}
Lattice methods have arrived. There are so many  lattice
calculations of different matrix elements that it is impossible
 to describe them
all, and in many cases the quality of the results is very high.
One can see plots showing extrapolations in lattice spacing which show
that the control of lattice spacing has become good enough to make
continuum predictions with small uncertainties.
Calculations with dynamical fermions and a small lattice spacing
are still nearly impossibly expensive to perform, and ``quenching'' remains
the dominant unknown in all lattice matrix element calculations.

There are two
major tasks facing lattice experts. I believe that all the people in
our field would agree that  the first problem
is to reduce the computational
burden, so that we can do more realistic simulations with smaller 
computer resources. I have illustrated several of the approaches people
are using to attack this problem. I believe that some of them have
been shown to be successful, and that ``improvement'' will continue
to improve.

There is a second question for lattice people, which I have not discussed,
but I will mention at the end: Is there a continuum phenomenology
of light hadron structure or confinement, which can be justified from
lattice simulations? The motivation for asking this question is that
there are many processes which cannot be  easily addressed via the
lattice, but for which a QCD prediction ought to exist.
For examples of such questions, see the talk of Bjorken
in this conference \cite{BJ}, or Shuryak's article\cite{SHURYAK}.
Few lattice people are thinking about this question.
Part of the lattice community spends its time looking for ``structure''
in Monte Carlo-generated configurations of gauge fields: instantons,
monopoles, $\dots$.
This effort is not part of the mainstream because the techniques either
involve gauge fixing (and so it is not clear whether what is being seen
is just an artifact of a particular gauge), or they involve arbitrary
decisions during the search (perform a certain number of processing
operations, no more, no less).
To answer this question requires new ideas and a controlled approach to
simulation data. Will an answer be found?

\section*{Acknowledgements}
I would like to thank 
M.~Alford,
C.~Bernard,
T.~Blum,
P.~Burrows,
S.~Gottlieb,
A.~Hasenfratz,
P.~Hasenfratz,
U.~Heller,
P.~Langacker,
P.~Lepage,
P.~Mackenzie,
J.~Negele,
F.~Niedermayer,
J.~Shigemitsu,
J.~Simone,
R.~Sommer,
R.~Sugar,
M.~Teper,
D.~Toussaint,
D.~Weingarten,
U.~Wiese,
and
M.~Wingate
for discussions, figures, and correspondence.
I would also like to thank Brian Greene, the organizer of the 1996
TASI summer school for asking me to lecture there: it means that I had
already written most of the material for these lectures well in advance
of the deadline!
I would also like to thank the Institute for Theoretical Physics
at the University of Bern for its hospitality, where these lectures
were written.
Finally, it was a pleasure to be able to visit SLAC again, talk to many
of my old friends, and present my subject to a wider audience.
This work was supported by the U.~S. Department of Energy.

\newcommand{\PL}[3]{{Phys. Lett.} {\bf #1} {(19#2)} #3}
\newcommand{\PR}[3]{{Phys. Rev.} {\bf #1} {(19#2)}  #3}
\newcommand{\NP}[3]{{Nucl. Phys.} {\bf #1} {(19#2)} #3}
\newcommand{\PRL}[3]{{Phys. Rev. Lett.} {\bf #1} {(19#2)} #3}
\newcommand{\PREPC}[3]{{Phys. Rep.} {\bf #1} {(19#2)}  #3}
\newcommand{\ZPHYS}[3]{{Z. Phys.} {\bf #1} {(19#2)} #3}
\newcommand{\ANN}[3]{{Ann. Phys. (N.Y.)} {\bf #1} {(19#2)} #3}
\newcommand{\HELV}[3]{{Helv. Phys. Acta} {\bf #1} {(19#2)} #3}
\newcommand{\NC}[3]{{Nuovo Cim.} {\bf #1} {(19#2)} #3}
\newcommand{\CMP}[3]{{Comm. Math. Phys.} {\bf #1} {(19#2)} #3}
\newcommand{\REVMP}[3]{{Rev. Mod. Phys.} {\bf #1} {(19#2)} #3}
\newcommand{\ADD}[3]{{\hspace{.1truecm}}{\bf #1} {(19#2)} #3}
\newcommand{\PA}[3] {{Physica} {\bf #1} {(19#2)} #3}
\newcommand{\JE}[3] {{JETP} {\bf #1} {(19#2)} #3}
\newcommand{\FS}[3] {{Nucl. Phys.} {\bf #1}{[FS#2]} {(19#2)} #3}

\end{document}